\documentclass[10pt,journal,compsoc]{IEEEtran}
\ifCLASSOPTIONcompsoc
  \usepackage[nocompress]{cite}
\else
  \usepackage{cite}
\fi
\ifCLASSINFOpdf
\else
\fi
\ifCLASSOPTIONcompsoc
  \usepackage[caption=false,font=footnotesize,labelfont=sf,textfont=sf]{subfig}
\else
  \usepackage[caption=false,font=footnotesize]{subfig}
\fi

\usepackage{amsmath,amssymb,amsfonts}
\usepackage{graphicx}
\usepackage{textcomp}
\usepackage{xcolor}
\usepackage{hyperref}
\usepackage{booktabs}
\usepackage{enumitem}
\usepackage{caption}
\usepackage{tabu}   

\begin{document}

\newcommand{\xh}[1]{\textcolor{red}{#1}}
\newcommand{\hm}[1]{\textcolor{blue}{#1}} %haichao
\newcommand{\ptb}[1]{\textcolor{green}{#1}} 
\newcommand{\TODO}[1]{\textcolor{red}{#1}}
\newcommand{\etal}{et al. }

\newcommand{\rvs}[1]{\textcolor{black}{#1}} %revision on 03/2023
\newcommand{\rvssec}[1]{\textcolor{black}{#1}} %revision on 08/2023

\title{\rvssec{Bimodal} Visualization of Industrial X-Ray and Neutron Computed Tomography Data}

\author{Xuan Huang,
    Haichao Miao,
    Hyojin Kim,
    Andrew Townsend,
    Kyle Champley,
    Joseph Tringe,
    Valerio Pascucci,
    Peer-Timo Bremer
    
\IEEEcompsocitemizethanks{\IEEEcompsocthanksitem Xuan Huan and Valerio Pascucci are with the SCI Institute, University of Utah.\protect\\
E-mail: xuanhuang@sci.utah.edu pascucci@sci.utah.edu
\IEEEcompsocthanksitem Haichao Miao and Hyojin Kim are with the Center for Applied Scientific Computing, Lawrence Livermore National Laboratory.\protect\\
E-mail: miao1@llnl.gov kim63@llnl.gov
% \IEEEcompsocthanksitem Hyojin Kim is with the Center for Applied Scientific Computing, Lawrence Livermore National Laboratory.\protect\\
% \TODO{E-mail:} 
\IEEEcompsocthanksitem Andrew Townsend, Kyle Champley and Joseph Tringe are with the Nondestructive Characterization Institute, Lawrence Livermore National Laboratory.\protect\\
E-mail: townsend10@llnl.gov champley1@llnl.gov tringe2@llnl.gov
\IEEEcompsocthanksitem Peer-Timo Bremer is with the SCI Institute, University of Utah and the Center for Applied Scientific Computing, Lawrence Livermore National Laboratory\protect\\
E-mail: bremer5@llnl.gov
}}

\IEEEtitleabstractindextext{%
\begin{abstract}
%why
Advanced manufacturing creates increasingly complex objects with material compositions that are often difficult to characterize by a single modality. 
Our \rvssec{ collaborating} domain scientists are going beyond traditional methods by employing both X-ray and neutron computed tomography to obtain complementary representations expected to better resolve material boundaries.
However, the use of two modalities creates its own challenges for visualization, requiring either complex adjustments of \rvssec{bimodal} transfer functions or the need for multiple views.  
%what
Together with experts in nondestructive evaluation, we designed a novel interactive \rvssec{bimodal}  visualization approach to create a combined view of the co-registered X-ray and neutron acquisitions of industrial objects. 
%how
%Using an automatic topological segmentation of the bivariate histogram of X-ray and neutron values as a starting point, the system provides a simple yet powerful interface to easily create, explore, and adjust a \rvssec{bimodal} visualization. 
Using an automatic topological segmentation of the bivariate histogram of X-ray and neutron values as a starting point, the system provides a simple yet \rvs{effective} interface to easily create, explore, and adjust a \rvssec{bimodal} visualization. 
We propose a widget with simple brushing interactions that enables the user to quickly correct the segmented histogram results.
%Furthermore, we introduce a new rendering mode that, instead of mapping opacity and color to histogram values, calculates opacity based on the distance to the nearest material boundary (in value space), and makes it possible to highlight otherwise difficult to extract features. 
Our semiautomated system enables domain experts to intuitively \rvs{explore} large \rvssec{bimodal} datasets without the need for either advanced segmentation algorithms or knowledge of visualization techniques.
%Our system enables domain experts to quickly analyze large \rvssec{bimodal} datasets without the need for either advanced segmentation algorithms or knowledge of visualization techniques.
%results and evaluation
We demonstrate our approach using synthetic examples, industrial phantom objects created to stress \rvssec{bimodal} scanning techniques, and real-world objects, and we discuss expert feedback.
\end{abstract}

}

\maketitle

\IEEEdisplaynontitleabstractindextext

% \fi
\IEEEpeerreviewmaketitle

\IEEEraisesectionheading{\section{Introduction}\label{sec:introduction}}

\IEEEPARstart{A}{dvanced} manufacturing techniques enable creation of increasingly complex parts and assemblies with internal structures, multiple materials, custom-designed impurities, and a host of other features.
However, to confidently utilize these capabilities in engines, buildings, or industrial facilities, resulting parts must be inspected and ultimately certified.
Furthermore, many of the more advanced manufacturing techniques are rapidly evolving, and the inspection of as-built parts becomes key to improving the process.
The challenge is that the additional degrees of freedom provided by additive manufacturing, i.e., 3D printing, often create internal features that cannot be observed with traditional \rvs{metrology} tools such as surface probes.
Instead, the field has moved to high-resolution X-ray computed tomography (CT) scans to enable volumetric analysis.
Although the X-ray scan provides the ability to observe internal structures, it creates new challenges in managing, analyzing, and visualizing large-scale volumes.
Members of our team are at the forefront of yet an additional challenge in this space where often a single scanning modality is insufficient.
%
% PTB: This makes it sound like we only deal with phantoms and this is not a real problem  
%
%They deliberately created phantoms, usually nested cylinders of different material types, to research the material characterization pipeline from the development of novel acquisition techniques to advanced visual analysis of combined X-ray and neutron tomography volumes. A major aspect is solving the identification of materials that have different signal responses in the two modalities.

%For example, X-ray computed tomography (CT), which is by far the most common tool, primarily distinguishes high density materials such as metals, but has a very limited ability to resolve features in organic materials such as plastics, according to our collaborating experts in nondestructive evaluation.
%Furthermore, sufficiently thick quantities of high density metals like tungsten often block X-rays entirely, resulting in impenetrable parts of an object.
As combinations of metals, plastic, and shielded parts are common in critical applications, developing approaches that can effectively characterize such objects is of significant interest to our collaborators. 
To address the limitations of one modality, additional modalities are used, and here we are primarily concerned with the addition of neutron-based tomography. Unlike the high-energy photons of an X-ray beam, neutrons interact only weakly with many materials \cite{banhart:2008}. 
As a result, neutron tomography does not suffer from the shielding effects of high-density materials.
Furthermore, unlike X-rays\rvssec{,} neutrons do interact relatively strongly with organic materials, such as plastics. 
Consequently, a neutron image can provide insights into materials invisible to X-rays as well as  information about otherwise shielded parts.
However, due to the limited power of available neutron sources, the spatial resolution of scans is lower overall in comparison to X-ray CT. 

Users are now faced with two \rvs{large-size} volumes corresponding to X-ray and neutron attenuation that must be analyzed jointly.
The ultimate goal for most industrial parts is a segmentation into materials for subsequent analysis.
However, the large volumes, \rvssec{bimodal} data, and common reconstruction artifacts, such as streaks or beam hardening, make any automatic segmentation challenging, and we currently have no accepted solution. 

To facilitate the process of initial data understanding, our goal is to create an exploratory visualization that can quickly provide an overview of an object, search for obvious flaws, and plan more detailed analysis steps. The result \rvssec{serves} as a solid basis for material scientists at an early stage of analysis, allowing them to make quick judgments on the scanning quality and the overall structure of the datasets before investigating  specialized analysis tasks.

Existing tools in the visualization community are largely based on linked view systems providing matching views of both channels or on complex multimodal transfer functions.
The former requires users to mentally assemble different \rvs{2D images}, which can be difficult especially for complex 3D assemblies.
The latter implies significant training and familiarity with the underlying visualization concepts that our typical user will not have.
Since neutron imaging remains an experimental capability, few dedicated solutions to these challenges exist, and current approaches to visualize \rvssec{bimodal} data in manufacturing are largely limited to a side-by-side view of orthogonal slices extracted through custom scripts, i.e., in matlab, which are not adequate for a meaningful exploration.

Here we present a new approach jointly designed by a team of nondestructive evaluation experts and visualization researchers to interactively explore \rvssec{bimodal} scans of industrial objects at an early stage of analysis. 
The system combines a semiautomatic topological segmentation of the bivariate histogram with an interactive approach to adjust the visual mapping from segments to visual properties and to explore the detected materials. 
%Finally, we include a \rvssec{bimodal} rendering component designed to highlight either material regions or material boundaries for both modalities.

The topological segmentation potentially coupled with localized thresholds using the relevance metric~\cite{Mascarenhas09, Bremer16camcs} provides a scalable preprocessing step, which creates an easy-to-explore space of hierarchical segmentations that the user can navigate by simply varying the number of segments shown.
The direct link to the joint histogram enables domain experts to quickly form hypotheses on the type of material that is segmented, and simple interactions such as assigning colors to materials and adjusting per-segment opacities provide a flexible interface to investigate hypothesises on the material composition. 

We identify our main contributions as follows:
 \begin{itemize}
        % novel segmentation
        % pre-segmentation
        \item 
        %A ready-to-use, semiautomated topological segmentation optionally enhanced with a relevance-based hierarchy on bivariate histograms for exploring multimaterial object scans with full flexibility. 
        %The computation of this topological hierarchy is fast and fully automatic, avoiding traditional labor-intensive manual segmentation, 
        %while the user remains in control to select the number and visual encoding of segments. 

        % \xh {removed the descriptive sentences, add how the particular choice of segmentation is better}
        \rvs{We propose the application of topological segmentation of the bivariate histogram for the exploration of multimaterial objects that are scanned with x-ray and neutron CT. This approach is enhanced by a relevance-based hierarchy, is fast to compute, and provides easy material decomposition overview of the multimaterial industrial objects.}
        %\rvs{A ready-to-use, semiautomated topological segmentation optionally enhanced with a relevance-based hierarchy on bivariate histograms for exploring multimaterial object scans with full flexibility. This topological hierarchy is fast to compute and carefully designed to provide easy material decomposition overview of the bivariate industrial objects.} 
        
        % helpful viewing modes
        \item %A novel approach to real-time rendering using  Intel's high-performance OSPRay engine~\cite{Wald16} for multimodal volumes that provides a combined view of co-registered X-ray and neutron volumes integrating the advantages of both modalities. (Section 5.1) 
        %\rvs{maybe "novel" is too strong}
        \rvs{A combined view of co-registered volumes integrating the advantages from both modalities, with straightforward color-material mapping and real-time rendering using Intel's high-performance OSPRay engine~\cite{Wald16}. (Section 5.1)}
        
        \item % flexibility
        \rvs{An interactive widget for the straightforward editing of the topological segmentation results allows us to adjust the number, color, and visibility of segments to avoid complicated transfer function design for each modality. We utilize the 2D space of the segmented bivariate histogram as an interface for merging, editing, and creating material segments.}

        %An interactive widget to facilitate easy adjustment of the number, color and opacity of segments to avoid complicated transfer function editing for each modality. Using the 2D space of the bivariate histogram as an interface enables easy visibility management of material segments. (Section 5.2)
        % wide range of data, flexible analysis environment
        % part of the rendering
        % closing of the intro
        \item 
        A thorough evaluation of two simulated datasets and three real-world acquisitions. We present three use cases proposed by experts from diverse backgrounds in the nondestructive evaluation field, ranging from CT reconstruction, analysis, and evaluation, to experimental validation. The result covers both objective comparison and subjective comments on the effectiveness of our system. (Section 6)
%
% PTB: This does not sound like a contribution
%
%        \item 
%        We require no prior knowledge of the data from the user and create a flexible exploratory %environment for the analysis of the material composition in industrial objects. 
\end{itemize}

\section{Related Work}
In this section, we look at previous work in \rvssec{bimodal} data visualization with a focus on volumetric data. In particular, we conduct a literature search of visualization for industrial CT. We also include approaches that focus on topology-based segmentation techniques, examine the use of transfer function design within the context, and discuss existing \rvssec{bimodal} visualization systems with similar case studies.

\subsection{\rvssec{Bimodal} Data Representation and Segmentation}
The terms \rvssec{bimodal} and bivariate have been used interchangeably to describe a dataset with more than one value at any sample point. One of the main reasons a \rvssec{bimodal} dataset is desired is that different modalities provide complementary information. The feature of interest is usually defined as a combination of values across different modalities. Thus, multimodality often causes problems in user interpretation, and an effective data representation is crucial for further analysis. The survey by Lawonn \etal~\cite{Lawonn17} examines various visualization techniques for CT data, but the solutions are spread across different medical subdomains, and there is no integrated framework for commonly used techniques.

Data segmentation is considered an important basic operation. Topological methods have been introduced to offer a mathematical abstraction~\cite{Correa11}, and Jankowai and Hotz developed an algorithmic solution to segment bivariant datasets through isosurfaces ~\cite{Jankowai21} for important features. These existing topology-based or isosurface-based approaches reduce the complexity in the target dataset by highlighting certain subdomains, but they pay little attention to a comprehensive initial overview. 

The potential of bivariate histogram segmentation has already been explored by LaManna \etal \cite{LaManna20}. The authors introduced a bivariate histogram method over two modalities to leverage material contrast in volumes~\cite{LaManna20}. Although LaManna \etal's work supports our decision to employ histogram segmentation, their approach remains a manual process of placing polygons inside the histogram \rvs{that outputs labeled binary volumes, but requires a typical active time of 15-30 minutes for visualization output}. Our work takes this idea further and includes a visualization system with semiautomated segmentation and interactive mapping of segments to visual properties. 

\rvs{Building on top of Kniss's \cite{Kniss02} multidimensional transfer function work, our segmentation method is also similar to the work of Wang et al.\cite{wang12} with automated Morse-complex based segmentation on the value vs gradient magnitude feature space. Although this system \rvssec{uses a similar} automated pipeline, the underlying 2D space asks for a visualization goal that does not work smoothly with \rvs{Morse-complexes}. With gradient magnitude as the second field, the materials to be identified do not correspond to the visually distinct arch regions, but rather circular structures at the bottom. Therefore, although the system overall provides an automated Morse-complex based solution like ours, the resulting segmentation is an indication but not a direct representation of different materials. In our system, the resulting segments align well with the target decomposition piece in the 2D space, and modifications can be made to match the material decomposition process more exactly.} 

\subsection{Rendering of \rvssec{Bimodal} Datasets }\label{related_work_multimodal}
% volume rendering is standard. Comparing to MIP and blending it's more comprehensive and can be speeded up by xyz. We rely on OSPRay cpu based ray-tracing for visual quality and interactive performance. 

\noindent \textbf{2D Projection:}
Since the final outputs of the visualization system are images, multiple modalities also have to be integrated into these final images in a reasonable way, such that the user is able to extract desired information.
One of the most straightforward multichannel blending schemes is through Maximum Intensity Projection (MIP). Fishman \etal have discussed the advantage of volume rendering versus MIP in CT angiography ~\cite{Fishman06} and concluded that MIP is best for fast demonstration or enhancement of certain organs, but lacks the ability to provide a clear definition of all structures such as soft tissues and muscles. 

On the other hand, volume rendering is also a standard solution due to its ability to produce a better visualization result in general. However, volume-rendering-based visualization usually relies on user mastery of rendering parameters and interpretation to reveal the feature of interests. Intermediate structures, such as isosurfaces or additional data analysis panels, are often necessary during the process and thus require additional mental effort.

\vspace{2mm}
\noindent \textbf{Data Fusion}: Another obvious direction is to combine all channels so that the standard scalar field volume rendering pipeline can be applied naturally. Therefore, the image fusion technique is widely used by industrial CT solutions to form a single scalar field \cite{Heinzl07} or to resolve a \rvssec{bimodal} image \cite{Mueller06}. Various metrics are discussed to blend multiple scalar fields ~\cite{Cai1999}, as well as the use of different intermixing schemes for direct volume rendering on GPU-based devices ~\cite{Schubert2011}.
Redesigning the color blending \rvs{function} also helps eliminate false additive colors~\cite{Morris02, Chuang09}, but qualitative analysis in the color space remains an open problem due to challenges in human perception.
In general, although the technique itself is valuable for producing high-quality fused datasets for specialized analysis conditions, fusion is performed as an extra merge operation outside the visualization system. Thus, such techniques raise the risk of information loss or bias and allow little flexibility in overview exploration.

\vspace{2mm}
\noindent \textbf{Direct Volume Rendering:}
Although 3D data rendering itself is well supported in modern graphics pipeline, multimodality poses additional challenges when targeting an interactive system.
Due to the \rvssec{bimodal} nature, the data size will \rvs{double for bivariate datasets}, and the multimodalities are considered to be only different RGBA channels that still require fusion. 
\rvs{Schubert and Scholl~\cite{Schubert2011} provide a data fusion solution with CUDA and GPU with interactive rate.} Ghosh proposes hardware-assisted volume rendering~\cite{Ghosh03} to speed up the process by efficient parallelization of multiple graphics hardware boards. Both techniques suffer from the inherent volume slicing artifacts, and a dual node PC cluster is used in the latter case because a single machine implementation does not perform well.

\rvssec{
The visualization system in this paper uses direct volume rendering with a CPU-based rendering engine \cite{Wald16}. The embedded high-performance ray-tracing method allows interactive rate on desktop machines for reasonable sizes scientific data and produces higher quality images. %Moreover, since we include no preproccessing or postprocessing filtering on the volume, together with the segmentation scheme in previous section users are able to expect a straightforward visualization result from the original dataset.
}

\subsection{\rvs{Transfer Function and Interface Design}}

Transfer function design directly affects user interaction with the goal of balancing generalization and precise user manipulation. Automated clustering-like solutions are well explored in 2D transfer function design for scalar field volumes ~\cite{wang12, Cai17}, but the solution for multivariate datasets is not straightforward \rvs{with dense value pairs, as in our cases}. 
To overcome the difficulties in single techniques, interpanel linking is widely adopted to reflect multiple changes interactively ~\cite{Buja91}. 

Kniss designed a multidimensional transfer function interface with dual-domain and classification widgets~\cite{Kniss02} that is much more complicated than the common scalar field transfer function widgets. Illustration-based approaches establish indirect mapping from multidimensional attribute space to color space through formulations, such as sets, numeric operations, and semantics~\cite{McCormick04, Woodring06, Rautek07}. Similarly, the idea of fiber surfaces introduces new representations of multivariate datasets as shown \rvssec{in the work of Athawale \etal \cite{Athawale23}}. Whereas these methods provide expressive visual abstractions for \rvssec{bimodal} datasets, the interpolated visual style elements may introduce ambiguity in material segmentation. \rvs{ SegMo \cite{NAGAI19} provides a straightforward Morse-complex based segmentation solution that is voxel-level precise but requires a heavy topological structure in volume space. In the direction of more specialized transfer \rvssec{functions}, Sereda \cite{Sereda06} discussed a semiautomatic hierarchical method based on the concept of the LH space boundary. The potential of dimension projection and parallel coordinates \cite{Guo12} is also explored. Lu and Shen also looked into using subspaces to reduce visual elements \cite{Lu17}. \rvssec{However,} the newly introduced ideas and complex \rvssec{panels} also pose challenges to users who are usually not in the field of visualization.}

\rvs{Finally,} Tzeng and Ma\cite{Tzeng04} showed a \rvssec{cluster-space} visual interface for preprocessed classified volume. We took a similar approach by including views of individual segments, but we provided a more flexible environment for users to make adjustments to the segmentation result interactively with the rendering.

This seminal work takes the segmentation from bivariate histogram, which is easily translated into color mapping on the bivariate volume attributes while hiding the complexity of rendering parameters. We also include a brushing and linking ~\cite{Keim02} functionality to enhance the connection between the histogram and volume rendering interface.
%With straightforward bivariate histogram segmentation, we developed an intuitive user interface consisting of only opacity sliders, color picker, and common 1D transfer function widgets. 

\subsection{\rvssec{Bimodal} Visualization Systems for Industrial CT}

A survey by Heinzl \etal ~\cite{Heinzl17} examined the field of material science visualization and proposed several high-level challenges. Integrated visualization systems are often required for generated material science data to investigate features of interest, and a tailored feature extracting pipeline also has to be embedded for further analysis. Schiwarth \etal ~\cite{Schiwarth18} described a workflow for a series of fiber-reinforced polymer X-ray computed tomography datasets. To show the analysis result for all features, such as fibers or voids, the system uses a combination of 3D rendering and 2D multivariate scalar plot matrix. ImNDT ~\cite{Gall21} pushed the interactivity of exploring multivariate material data into an immersive virtual reality experience, by showing an overview with a grid-like displacement system so that the user can subdivide the domain spatially.

Our work avoids the high-dimensional charts or any additional data interpretation UI panel in that the bivariate histogram serves both as the result of segmentation and as color-mapping on the transfer function interface. \rvssec{Together with a multi-resolution data loading mechanism, by using an high-quality ray-tracing engine, we  utilized the built-in state-of-the-art volume rendering data structures and algorithms that are parallelized and optimized for modern instruction sets, and thus enables an interactive rate, higher fidelity volume rendering for all datasets we use (see \autoref{tab:datasets}). Moreover, both the segmentation and the rendering components are designed in a modularized way such that other algorithms can be easily swapped into the pipeline for even greater flexibility in application development.}

\section{Background}

Our domain experts are scientists focusing on the development of novel methods for the characterization of advanced materials and industrial objects. They have carefully constructed a range of phantom datasets to stress test their approach, which we use in this work. The phantoms are multimaterial objects, containing small features that stress the single modality, shown in \autoref{tab:datasets}. 
Furthermore, as manufacturing these test objects and acquiring data with different parameters are time-consuming and expensive, our experts rely on physics-based simulations~\cite{champley:2016:ltt} to create additional data as the \rvssec{scanning systems} are being developed. Physics-based simulations have the additional advantage that different setups can be easily explored, such as varying X-ray and neutron energies to plan for hardware changes or future systems.

X-ray imaging, although appreciated for its high spatial resolution, struggles with high-density metals, such as tungsten. As a result, important features could be hidden behind these materials, leaving this modality vulnerable to overlooking crucial defects. Neutron imaging, on the other hand, has a similar resolution everywhere, but a lower resolution overall \rvssec{\cite{Vontobel06}}. The combination of these two modalities promises a vast improvement in the ability to detect features that are otherwise difficult to identify. 
%Figure \autoref{fig:jh2_demo:b} shows the X-ray of an object with various nested cylinders consisting of different materials, as shown in the ground truth data in Figure \autoref{fig:jh2_demo:a}. We can see that the center is not resolved due to the shielding by the tungsten, while neutron CT, as shown in Figure \autoref{fig:jh2_demo:c}, depicts the small inclusions in the center, but has the same signal response for aluminium and plastic. 

The complementary nature of these two modalities motivated our collaborators to construct a scanner that is able to acquire X-ray and neutron volumes simultaneously, where the X-ray and neutron sources are offset by a fixed angle. As a result, the data will not require computational registration. Our collaborators primarily try to identify the materials and their boundaries. \rvs{Currently, this task relies on a side-by-side comparison of orthogonal slices from the two modalities extracted through scripts or simple interfaces. } 

% \subsection{X-ray and Neutron CT Scanning for NDE}
% \rvs{
%  what background is needed to understand our approach
% \begin{itemize}
%     \item CT inspection, how it's currently done
%     \item difference between x-ray and neutron
%     \item why is merging the two modalities important 
%     \item how is this different from data fusion 
%     \item details about data
%     \item what kind of objects
%     \item why segmentation with morse-smale complex and not reeb space or any other tool
% \end{itemize}
% }

\subsection{Bivariate X-Ray and Neutron Data }
We use seven X-ray and neutron CT datasets: JH2A ,JH2B, XR05A, XR05B, Synthetic Cylinder, battery and meteorite. JH2A, JH2B, and XR05A are \rvs{physics-based} simulations of nested cylinders with \rvs{a small volume of} material inclusions in the center of the object. Whereas JH2A and XR05A contain similar rod-shaped inclusions in the center with varying sizes, the JH2B dataset contains even smaller inclusions. 

The Synthetic Cylinder is created without physics-based simulation to clearly demonstrate the advantage of bivariate histogram segmentation, and all its materials are concentric cylinders in space. \rvssec{All datasets have matching sizes on both modalities and are co-registered. All histograms and thus the following segmentations are computed on full resolution. The last two datasets, Battery and Meteroite, are downsampled to 1/4 on each axis in the rendering process for desktop performance (Section \ref{section_Multimodal_vis}).}
To demonstrate our interactive approach to actual acquisition, we also include XR05B and battery, which are scanned from constructed objects, and a meteorite dataset that we analyze without any previous knowledge.

\begin{table}[]
    \centering
    \caption{The datasets used in this work. The phantoms are created by our collaborating experts to test acquisition, analysis, and characterization of materials using X-ray and neutron imaging. The phantoms are created using a \rvs{physics-based} simulation \cite{champley:2016:ltt}}
    \label{tab:datasets}
    \begin{tabular}{lcr}
        \toprule
        Phantom object & Type & Size \\
        \midrule
        JH2A & Physics-Based Simulation & 2x128x128x128\\
        JH2B & Physics-Based Simulation & 2x1000x1000x128\\
        XR05A & Physics-Based Simulation & 2x512x512x512\\
        XR05B & Acquisition & 2x256x256x256\\
        Synthetic Cylinder & Simulation & 2x512x512x512\\
        Battery & Acquisition & 
        \rvssec{2x2159x2159x2559}\\
		Meteorite & Acquisition & \rvssec{2x1999x1999x1549}\\
        \bottomrule
    \end{tabular}
\end{table}

% \begin{table}[b]
%     \centering
%     \caption{The datasets used in this work. The phantoms are created by our collaborating experts to test acquisition, analysis, and characterization of materials using X-ray and neutron imaging. The phantoms are created using a \rvs{physics-based} simulation \cite{champley:2016:ltt}}
%     \label{tab:datasets}
%     \begin{tabular}{lcr}
%         \toprule
%         Phantom object & Type & Size \\
%         \midrule
%         JH2A & Physics-based Simulation & 2x128x128x128\\
%         JH2B & Physics-based Simulation & 2x1000x1000x128\\
%         XR05A & Physics-based Simulation & 2x512x512x512\\
%         XR05B & Acquisition & 2x256x256x256\\
%         Synthetic Cylinder & Simulation & 2x512x512x512\\
%         Battery & Acquisition & 2x540x540x640\\
% 		Meteorite & Acquisition & 2x512x512x512\\
%         \bottomrule
%     \end{tabular}
% \end{table}

\begin{figure}[htb]
\centering
    \includegraphics[width=0.5
\linewidth]
{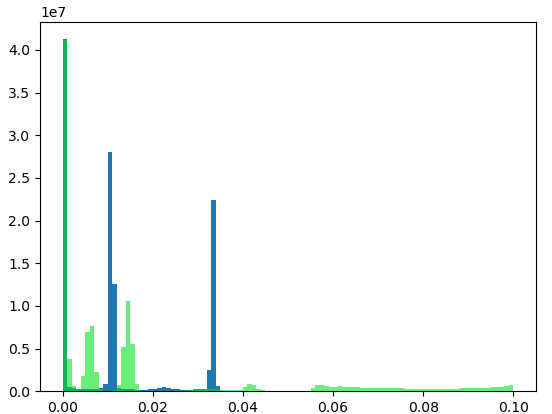}
  \caption{\rvssec{ The 1D histogram of the X-ray (blue) and neutron channel (green) of JH2B, which has five materials. The distinct peaks differ greatly, showing that one modality is inherently insufficient to capture all materials. In the X-ray, we can see three distinct peaks, whereas the neuron has three lower density peaks and a wider range of values that are spread out without any significant additional peak. }
  }
  \label{fig:demo_bivariate}
\end{figure}
In \autoref{fig:demo_bivariate}, we can see two histograms of the two modalities of the JH2B dataset, which consists of five materials. The X-ray and neutron clearly react differently to the materials, resulting in distinct peaks over the 1D histogram. As demonstrated by the number of peaks, neither single modality can capture all the material in the phantom object. 
%In addition, we also depict the corresponding slice views in \autoref{fig:jh2_demo}. As we can see, 
The X-ray provides an overall high-quality capture while losing precision at the interior after hitting high-density metal. The neutron behaves consistently with less detail, but is able to capture the internal structure.
We demonstrate our method with JH2A and JH2B in \autoref{section_Multimodal_vis} and \autoref{section_renderer}. All datasets are discussed in more detail in \autoref{section_result} with the domain experts.

\section{Multimodal Visualization of X-ray and Neutron CT}\label{section_Multimodal_vis}

\begin{figure*}[htb]
\centering
  \includegraphics[width=0.9\textwidth]{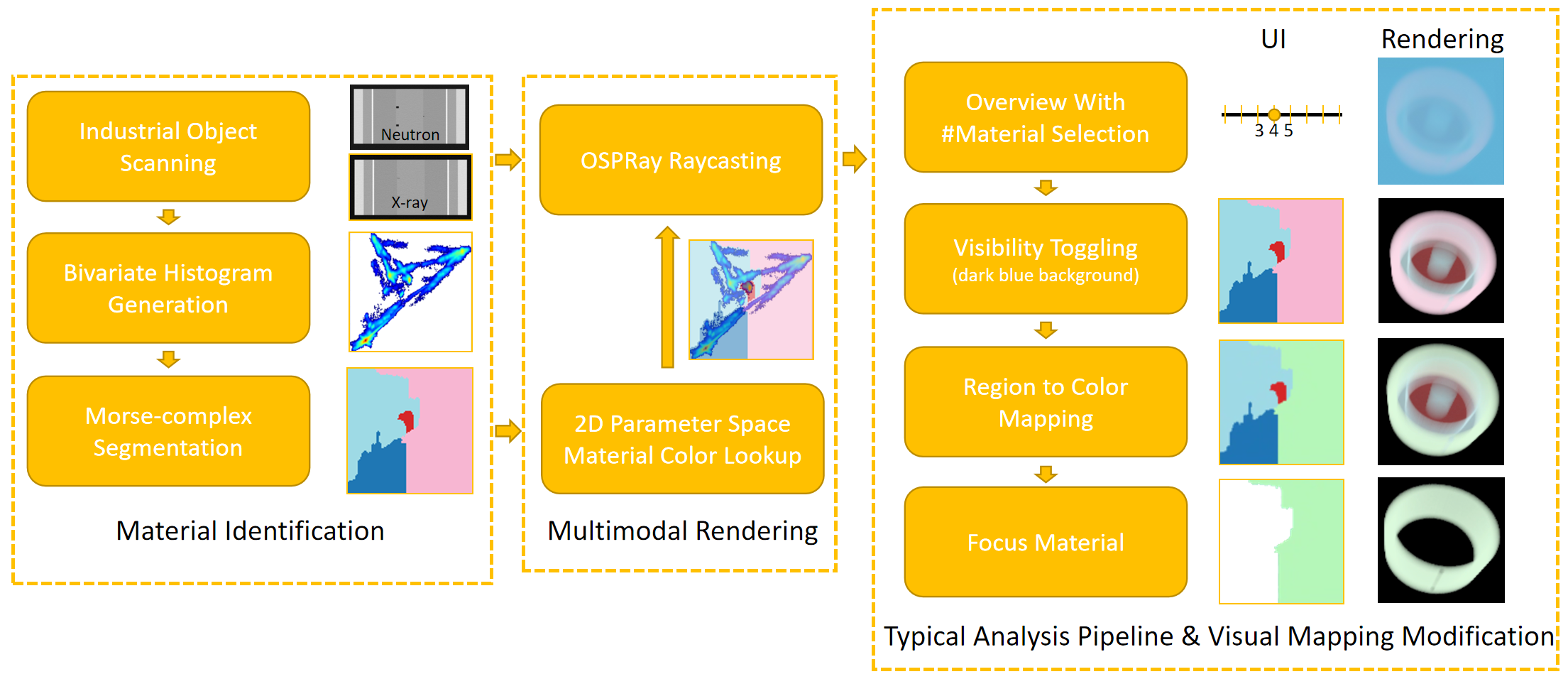}
  \caption{\label{fig:overview} Pipeline overview. We load the \rvssec{X-ray} and neutron data at full resolution, generate the bivariate histogram, and compute the Morse-complex segmentation. Both the segmentation and the dataset itself are set as input for the renderer. During the ray casting process, when a voxel is queried it will be sampled for both modalities, resulting in a pair of sampled values. The 2D vector corresponds to the 2D position in the bivariate histogram. The RGB values are determined by the same coordinate lookup in the segmentation image, which is then modifiable through the \rvssec{application}'s built-in GUI.}
  \vspace{-3mm}
\end{figure*}
We work with domain experts in nondestructive evaluation and design the visualization system with the goal of an efficient multimodal data analysis. The main idea is to provide a quick overview of features from both X-ray and neutron CT for the scientists without going into technical details in the field of visualization. Although experts traditionally inspect the difference in density values, their overall goal is to understand the material compositions. Hence, our work offers both an effective overview with minimum user interaction as well as the flexible inspection of individual materials. 

The pipeline is summarized as follows (\autoref{fig:overview}): 
The material identification is semi-automated. Our approach first creates the bivariate histogram and a topological hierarchy. The user can specify the number of segments, which are then assigned color and opacity. For analysis, the user can toggle the visibility of each segment through the interaction panel. \rvssec{Since some datasets are too large to fit into memory, we compute the histogram on the full-resolution volume but utilize the multiresolution approach by Kumar et al. \cite{Kumar10} and the OpenVisus\cite{visus} framework to downsample the data for rendering.} Our multimodal renderer looks up the color based on the localization of each value pair in the histogram and visualizes each segment with the respective color. 

With a Morse-complex based segmentation on the bivariate histogram formed by both modalities, we enable the user to oversegment the bivariate histogram from the two volumes into regions that roughly correspond to object materials. The outputs from this step are 2D segmentation images with color assignment of various number of segments. All images will then be taken as input into the renderer as lookup tables. Material colors can be adjusted by modifying the segmentation image on a simple interface. 
%Furthermore, since identifying material boundaries is usually one of the most important visualization goals in X-ray and neutron CT data, we also design a boundary mode to emphasis the structure. This mode is achieved by rendering a precomputed third channel maximum gradient magnitude volume to include all possible boundaries from both modalities. 

%We will be using JH2A throughout section 3 and 4 for demonstration. The results will be shown with industrial CT datasets and reports from domain experts using our system.

%\hm{quick summary of entire pipeline
%  \begin{itemize}
%\item we want to make TF specification easier. the idea is we oversegment the two volumes into regions that roughly correspond to materials and then assign these regions individual TF in bivariate histogram map.
%    \item we calculate the bivariate histogram of x-ray and neutron values. we use ms complex segmentation to get segments. we compute a distance transform based on the boundaries of segments. 
%    \item we assign for each segment a color and a mapping for
%      distance to opacity. we use pre-specified distance-to-opacity
%      functions, either material focused or boundary focused. we use a
%      ospray for rendering
%      \end{itemize}
%}

%\hm{highlight left fully automatic in left column.
%discuss in the end that we could do a better segmentation, but our users are not able to 
%}

\subsection{Domain Goals in Nondestructive Evaluation}
We conducted our work as part of a large project aimed at developing new methods for the nondestructive evaluation of industrial objects. An important component is the use of multiple modalities that produce complementary images of the same objects \rvs{because, as shown in \autoref{fig:demo_bivariate}, a single modality is insufficient to capture all materials in the object. Over the course of a year, we met weekly to discuss challenges and refine goals for the multimodal analysis. We were tasked with developing an effective method to examine X-ray and neutron data simultaneously without the need to mentally match multiple views of slices, so that the scientists  are able to gain initial yet comprehensive knowledge of the acquired datasets to understand material compositions. }

% Xuan: remove the details here
%Over the course of a year, we met weekly to discuss challenges and refine goals for the multimodal analysis. These meetings ensured close collaboration with the experts, by allowing us to ask questions and receive regular feedback. A basic assumption of the field, as shown in \autoref{fig:demo_bivariate}, is that a single modality is insufficient to capture all materials in the object,
%and the first challenge is that the materials in both modalities are represented with different intensity ranges. This multi-modality nature makes it difficult to create a spatial correspondence when using a side-by-side view, especially when looking at slices. One obvious consideration is to segment the data, but the segmentation result would not allow experts to understand which material region is associated with which intensity regions in the data. Another viable solution would be to fuse the data. Although this is considered a final solution to the two-volume problem, developing a fusion approach would first require an exploratory method that allows experts to change the components to be fused and determine how each modality should eventually contribute to the final fused image. Therefore, interactive analysis is the key to developing advanced fusion methods that would follow this initial observation stage. 

%\rvs{R1 suggest that the project background is not too relevant. I do see the intention here duplicates with motivation in some sense. Cut this down and incorporate into introduction or motivation eariler.}

From the above-mentioned considerations and description of challenges, we identify the following goals in terms of visualization and analysis:
\begin{itemize}
    \item \rvs{Show} an efficient overview of materials in industrial objects that are simultaneously scanned with X-ray and neutron CT.
    \item \rvs{Avoid} labor-intensive operations such as adjusting transfer functions or manual segmentation.
    \item Leverage existing views that are already familiar to the experts.
    \item \rvs{Enable} interactive exploration and the ability to see the shapes of individual material regions.
    \item Keep user interactions \rvssec{at} a minimum while providing flexibility in segmentation and visibility parameters updates. 
    
\end{itemize}

%Thus the main goal is to create a visualization that clearly distinguish different materials from each other, with a combined view that incorporates information from both modalities. 

%The result is this bivariate histogram based automated visualization system. In this section we show the advantage of our design in three industrial datasets, JH2, xr05 and a meteorite dataset, in that our tool has better image quality, fully automated material segmentation and requires minimum user input for a wide range of bivariant CT datasets.
%The common visualization analysis goal includes overview of both modality, identify different materials and their spacial correlations. 
%\hm{
%\begin{itemize}
%    \item we collab with them, weekly meetings, part of the bigger project, yadda yadda
%    \item what's the overall goal
%    \item Goal: be able to inspect the two modalities together
%    \item G: be able to distinguish materials and their boundaries
%    \item G: don't want to manually segment and draw material boundaries
%    \item G: material boundaries are fuzzy, be able to see that
%    \item G: be able to change material to visual properties assignment in an easy way (without fiddling around too much with transfer functions)
%    \item G: make regions transparent
%    \item not goal: exact segmentation
%\end{itemize}
%}
% \subsection{CT Data}
% CT data description. Raw data format we are using.
\begin{figure}[!htb]
\centering
    \subfloat[][Ground truth]{\includegraphics[width=0.32\columnwidth]{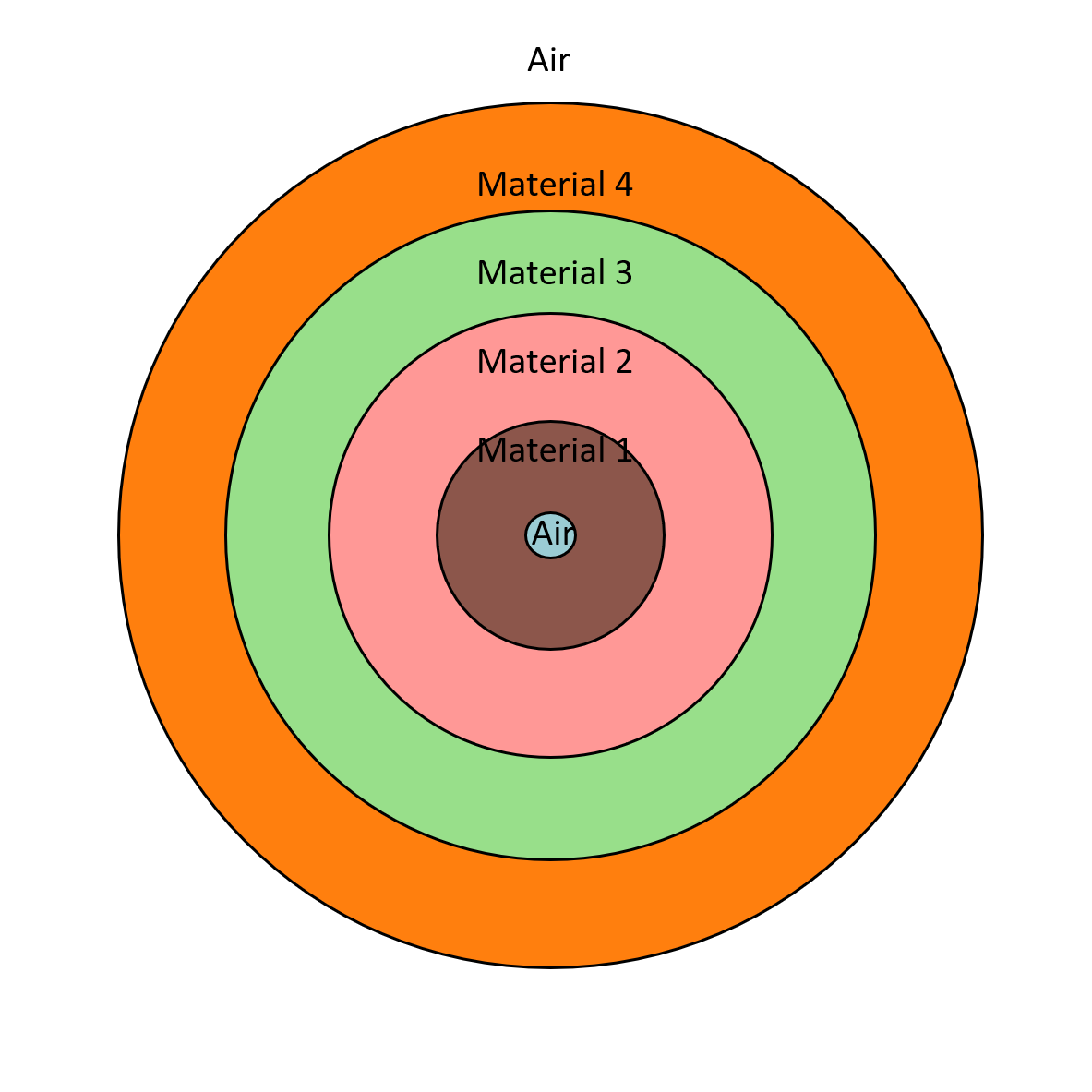}\label{fig:synthetic:a}}
    \subfloat[][X-ray]{\includegraphics[width=0.32\columnwidth]{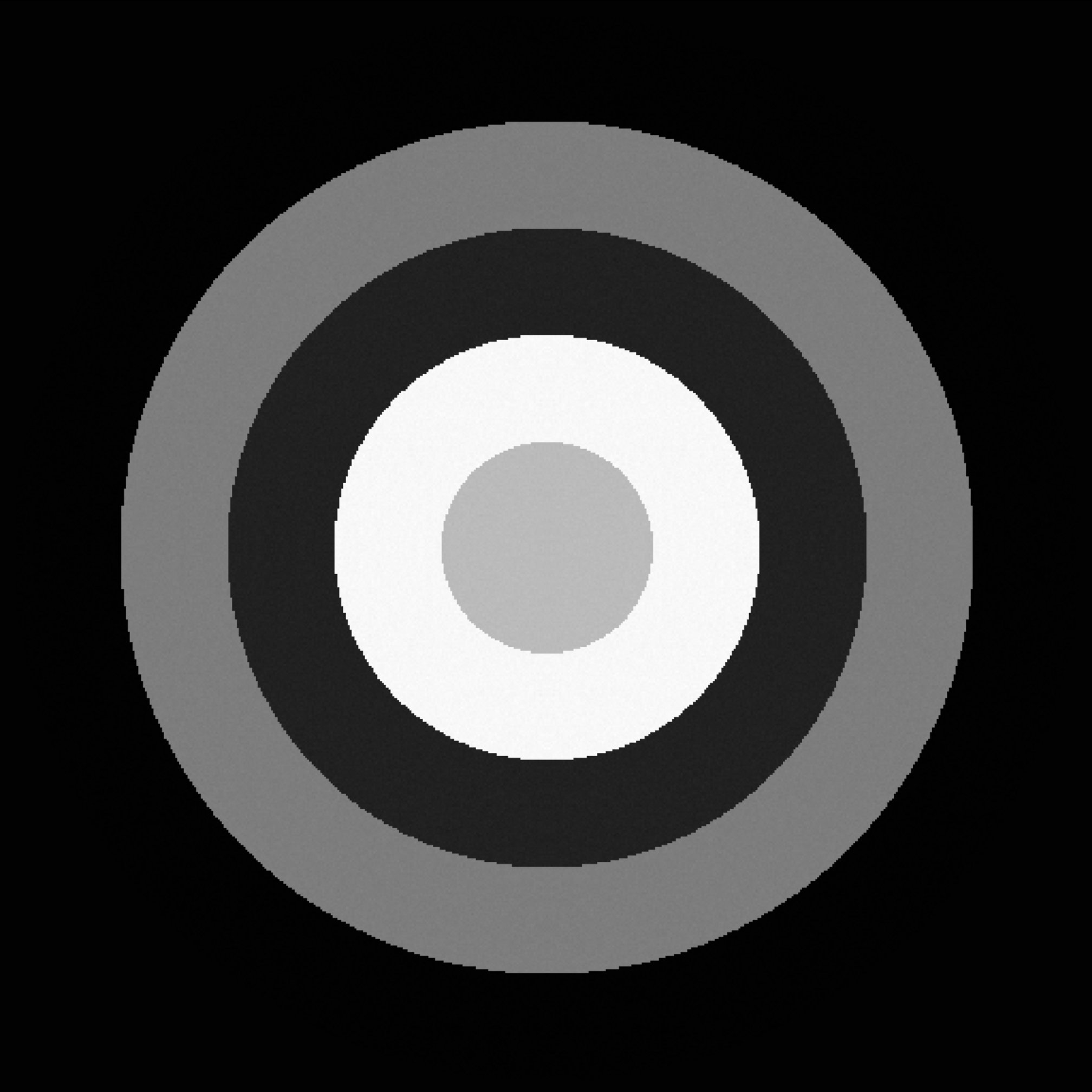}\label{fig:synthetic:b}}
    \subfloat[][Neutron]{\includegraphics[width=0.32\columnwidth]{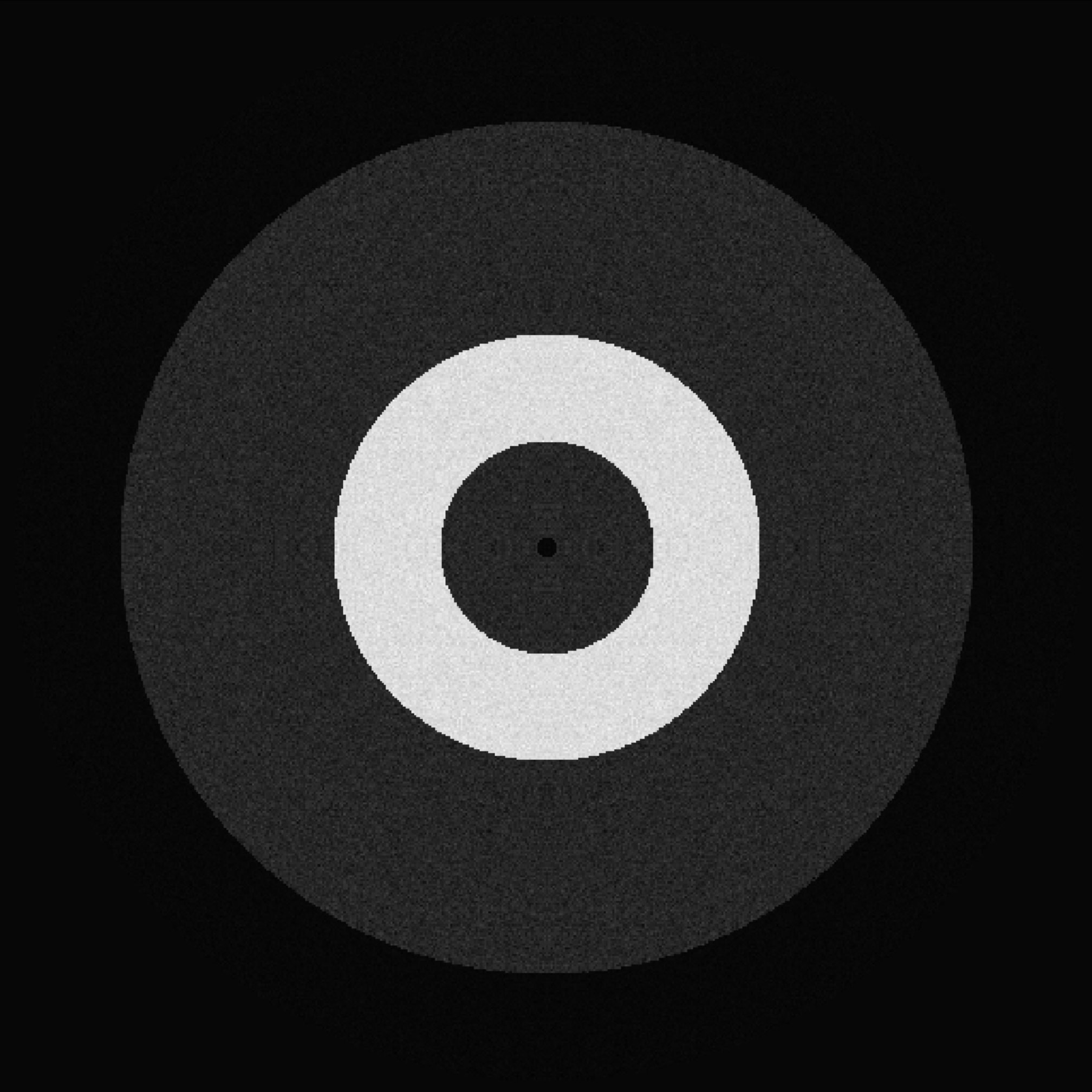}\label{fig:synthetic:c}}
    \\
    \subfloat[][Left: Histogram with rainbow colormap. Right: Segmentation result]{\includegraphics[width=1\columnwidth]{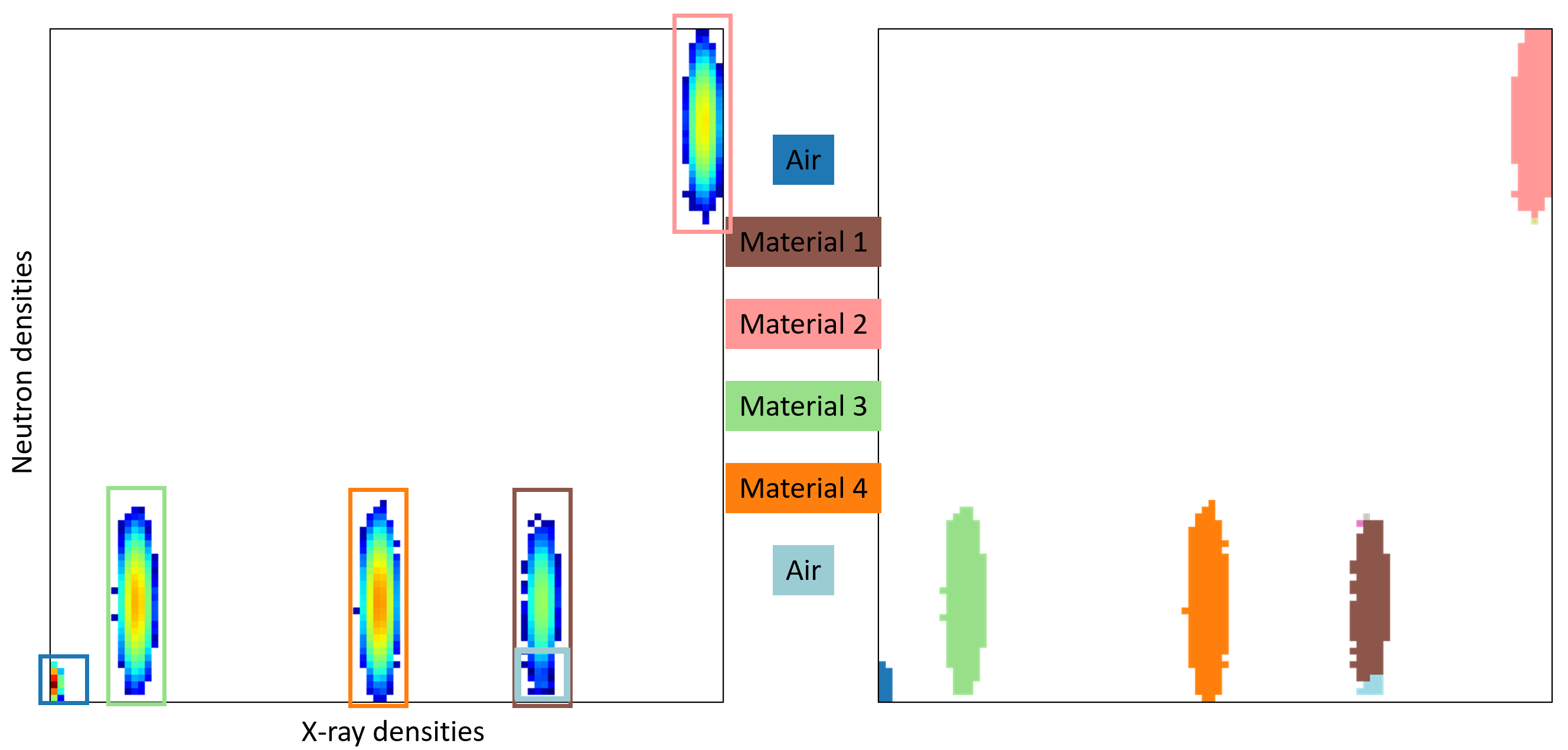}\label{fig:synthetic:d}}
    \caption{\label{fig:synthetic}
            %A Test example: The Synthetic Cylinder. (a) Shows the ground truth, (b) and (c) show a slice of the X-ray and neutron respectively. The X-ray fails to resolve the internal details due to the pink material, which is a synthetic high-density material. Although the neutron CT fails to depict the difference between the orange and green materials, it captures the small inclusions at the center. (d) depicts the bivariate histogram colored by its density value, as well as the segmentation from the Morse-complex segmentation driven by relevance. This dataset is carefully simulated to contain 4 concentric cylinders of different combinations of values from the X-ray and the neutron channel. The materials and backgrounds are well separated into 4+2 clusters on the bivariate histogram on the left, and the information is successfully captured in the Morse-complex segmentation on the right.
            \rvssec{
            A Test example: The Simulated Synthetic Cylinder that contains four concentric cylinders of different combinations of values from the X-ray and the neutron channel. (a) Shows the ground truth, (b) and (c) show a slice of the two channels respectively, with X-ray missing the innermost hole and Neutron falsely mixing orange and green materials. (d) Shows that the materials and backgrounds are well separated into 4+2 clusters on the bivariate histogram on the left. The information is successfully captured in the Morse-complex segmentation on the right.}
            }
            \vspace{-3mm}
\end{figure}

\subsection{Bivariate Histogram Generation}
\rvs{1D histograms are commonly employed by the experts to understand the material compositions without spatial information. The bivariate histogram has already been shown to be effective for identifying distinct materials from X-ray and neutron data \cite{LaManna20}}. 
\rvssec{Another advantage in utilizing the value frequency in the bivariate histogram for segmentation and the inherent binning of the values is that it also accounts for the previously mentioned artefacts in the acquisition process. If the value pairs are directly taken as input, the subsequent morse-complex segmentation would result in false positive segments as a result. Finally, we also consider the familiarity of our collaborators with bivariate histograms for material identification and this approach enables us for a seamless adoption to their existing workflow.}

%Hence, we use the segmented bivariate histogram as the basis for interaction, such as selection of material segments and managing visibility.  

The bivariate histogram provides the underlying data for the segmentation. The materials appear as \rvs{connected pixel areas} formed by different value combinations of the two modalities, but not all materials will necessarily result in high peaks. As a 2D domain, the histogram consists of, for example, 100x100 bins that provide us with a quick view of material clusters. The peak in the histogram will likely correspond to a material type, as in the 1D case. We have seen in \autoref{fig:demo_bivariate} that each 1D histogram has fewer material peaks than the actual number of materials. \rvs{It is} easy to see in \autoref{fig:synthetic} that the X-ray 1D histogram on the horizontal axis has brown and light blue materials as similar values, which make those indistinguishable on purely X-ray, whereas the neutron 1D histogram on the vertical axis will not separate green, orange, and brown materials. However, in the bivariate histogram, we can see the distinct material peaks in the 2D domain. Figure \autoref{fig:synthetic:d} shows five clusters that, in a simplified assumption, correspond to the materials. The background peak on the bottom left is naturally the highest.

\subsection{Morse-Complex Segmentation}
\label{sec:morsecomplex}
\begin{figure*}[!htb]
\centering
    \subfloat[][]{\includegraphics[width=0.32\linewidth]{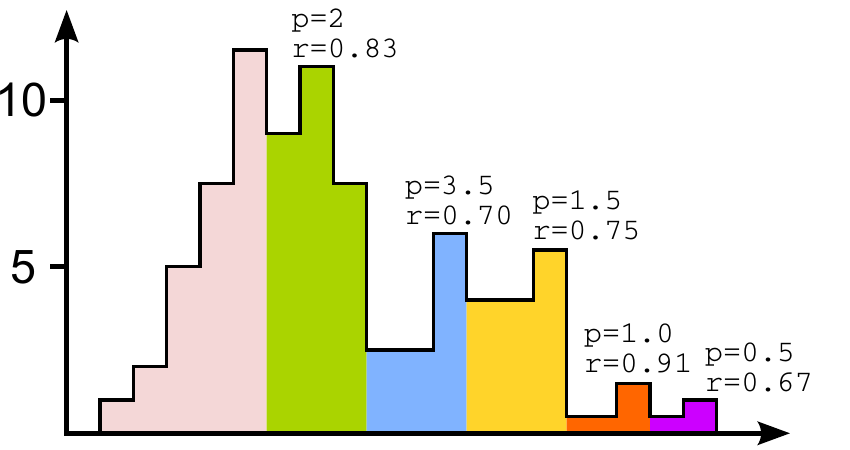}}\hfill
    \subfloat[][]{\includegraphics[width=0.32\linewidth]{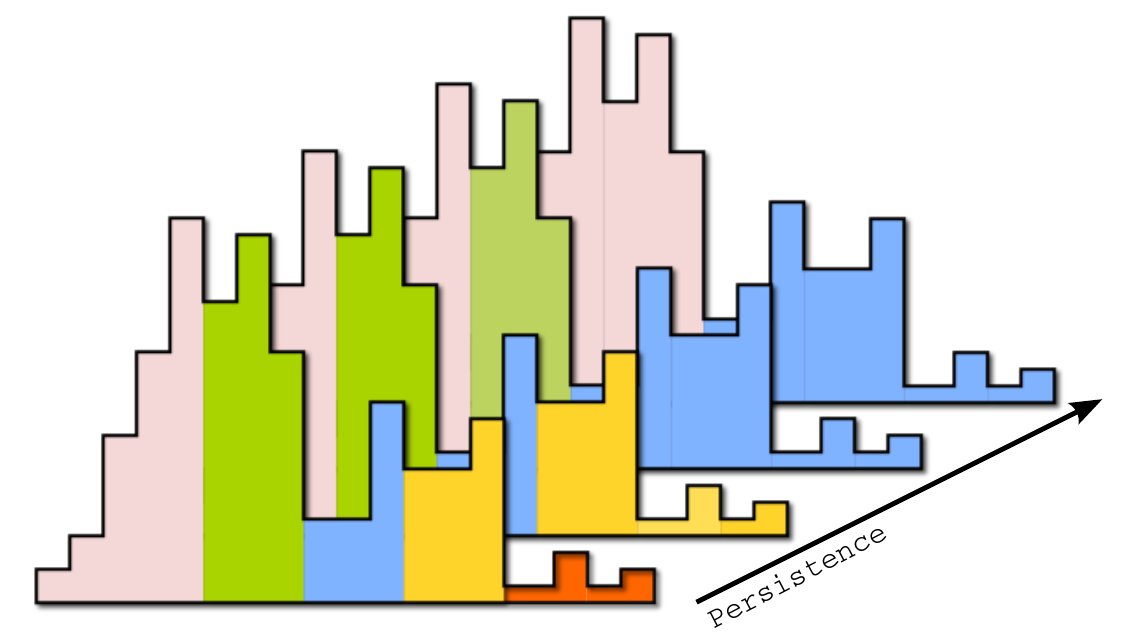}}\hfill
    \subfloat[][]{\includegraphics[width=0.32\linewidth]{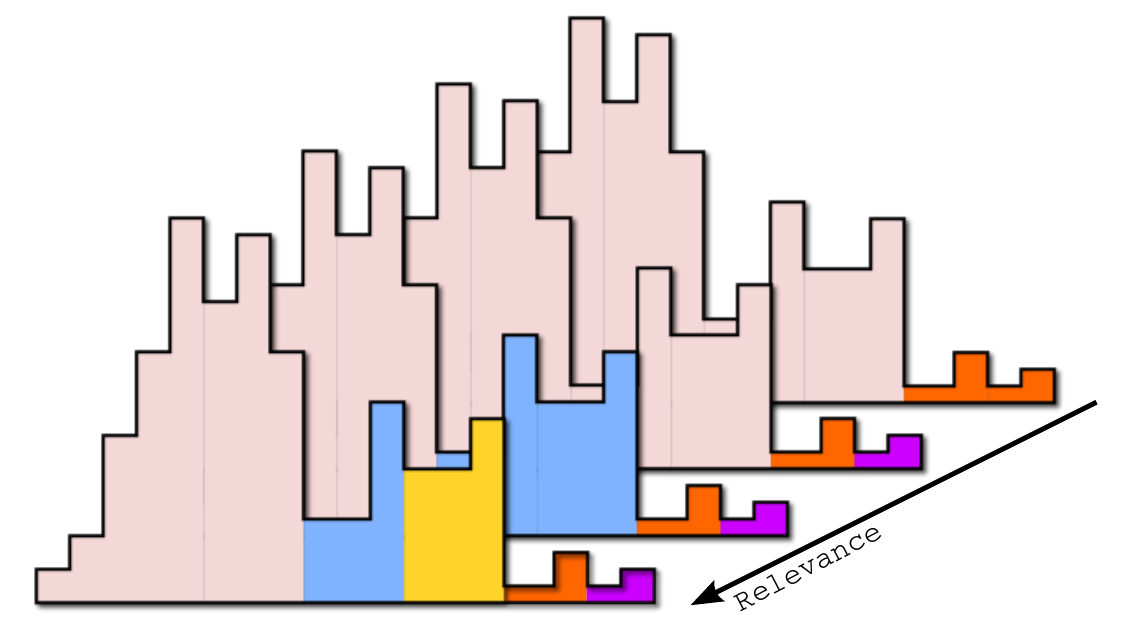}}
  \caption{Persistence vs. relevance simplification in a 1D histogram. (a) Original histogram segmented into modes indicated by color with the relevant persistence and relevance values of the local maxima indicated. (b) Hierarchical simplification by increasing persistence, which prioritizes high value peaks corresponding to common values. (c) Hierarchical simplification by decreasing relevance, which prioritizes distinct peaks when compared to their local neighborhood and preserves rare values.}
  \label{fig:persist_relevance_overview}
  \vspace{-3mm}
\end{figure*}

\begin{figure}[htb]
\centering
    \subfloat[][Histogram]{\includegraphics[width=0.32\columnwidth]{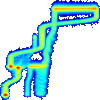}}\hfill
    \subfloat[][Persistence]{\includegraphics[width=0.32\columnwidth]{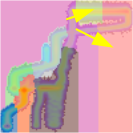}\label{fig:persist_relevance:persist}}\hfill
    \subfloat[][Relevance]{\includegraphics[width=0.32\columnwidth]{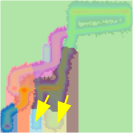}\label{fig:persist_relevance:relevance}}
    
    \subfloat[][Ground truth]{\includegraphics[height=0.32\columnwidth]{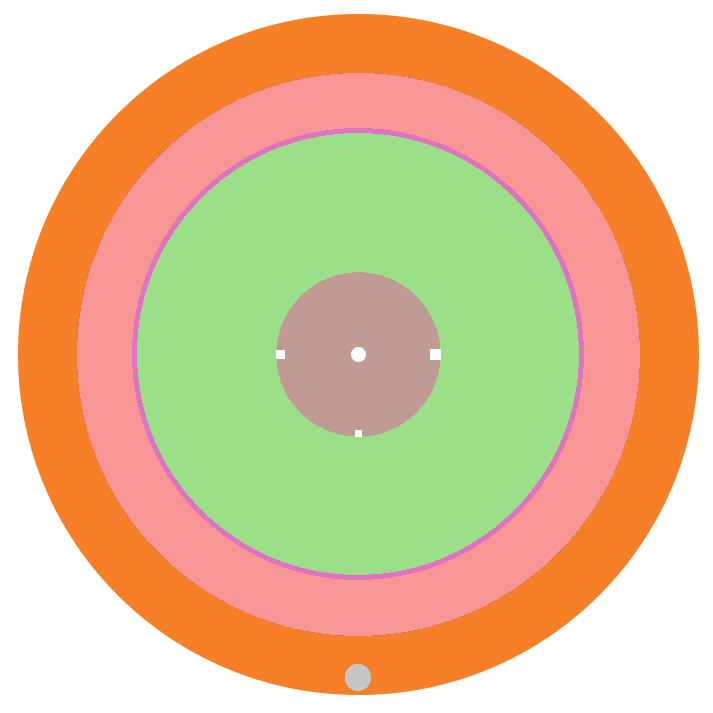}}\hfill
    \subfloat[][Persistence volume]{\includegraphics[height=0.32\columnwidth]{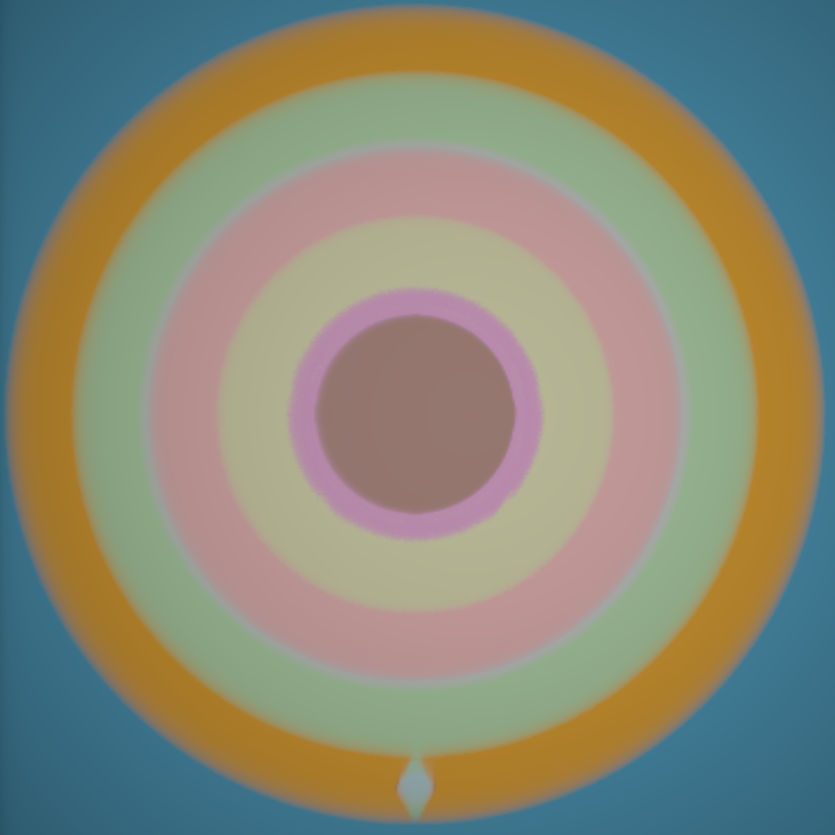}\label{fig:persist_relevance:persist_vol}}\hfill
    \subfloat[][Relevance volume]{\includegraphics[height=0.32\columnwidth]{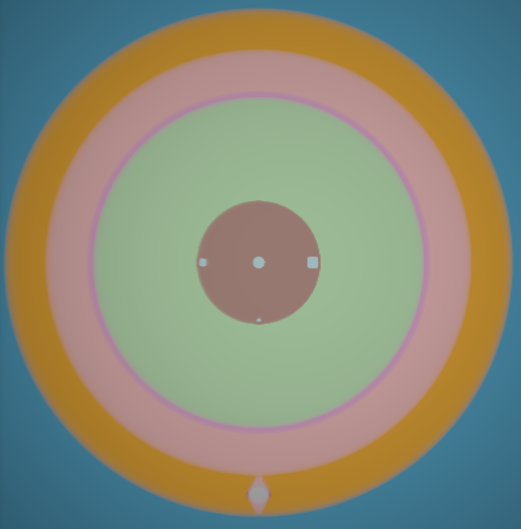}\label{fig:persist_relevance:relevance_vol}}
  \caption{%\label{fig:persist_relevance}
           \rvssec{JH2B} bivariate histogram (a) segmented with persistence (b) and relevance (c). With persistence, \rvs{the purple color segment in (b) tries to further split the highest value peak into yellow and pink segments indicated by arrows, whereas the relevance imposes a better segment-material correlation by identifying a possible new region around the lower value area (brown, light blue, and yellow in (c)).} (d)-(f) shows the ground truth illustration  and the volume result of segmenting by persistence vs by relevance. The persistence scheme tends to segment out the bigger volume with smooth value variation, whereas relevance prioritized the smaller pieces with more distinct value peaks that are more likely corresponding to individual materials. } 
           \vspace{-2mm}
\end{figure}

\begin{figure}[htb]
\centering
    \subfloat[][Manual]{\includegraphics[width=0.23\columnwidth]{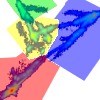}}\hfill
    \subfloat[][4 segments]{\includegraphics[width=0.23\columnwidth]{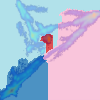}}\hfill\subfloat[][6 segments]{\includegraphics[width=0.23\columnwidth]{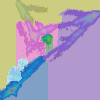}}\hfill
    \subfloat[][8 segments]{\includegraphics[width=0.23\columnwidth]{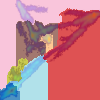}}\\
     %\subfloat[][2 segments]{\includegraphics[width=0.32\columnwidth]{img/_100_100_2_segs.png}}\hfill
    %\subfloat[][4 segments]{\includegraphics[width=0.32\columnwidth]{img/_100_100_4_segs.png}}\hfill
    %\subfloat[][7 segments]{\includegraphics[width=0.32\columnwidth]{img/_100_100_7_segs.png}}
  \caption{\label{fig:manual }
           The manual polygonal segmentation often conducted by domain experts (a) and our automated segmentation (b)-(d). As we can see, our segmentation provides detailed boundaries in comparison to the linear segments in the manual method. Although our tool did not pick up the exact same segmentation as desired when specifying four segments, with a higher number of segments it marked all important materials with precise boundaries, assigned colors from a qualitative color map.}
\vspace{-2mm}
\end{figure}

%\vspace{-10mm}
\subsubsection{\rvs{The Base Algorithm}}
Similar to other approaches that have used topology-related approaches to segment bivariate histograms ~\cite{Kotava13,vazquez07,Cai17}, the key insight is that, at least in our use case, such histograms are formed through a superposition of blurred Dirac functions, one for each identifiable material, including the surrounding air, i.e., the background.
In a perfectly resolved scan, with uniform materials, and without noise or artifacts, each material would be identified through a unique neutron-X-ray pair of values and be confined to a single bucket of the histogram. In practice, noise in both the X-ray and neutron sources as well as in the detector, assumptions in the underconstrained reconstruction problem, material variations, partial volume artifacts, etc., cause these Dirac peaks to be blurred into a more continuous response.
Computing the Morse-complex, also called the watershed, of this response function identifies all local peaks and their corresponding region of influence and thus splits the contributions of the various materials \rvs{as well as} possible without assuming further knowledge of the system, i.e., the number and types of materials present or the total amount of material in an object.

Here, we use a simple steepest gradient style algorithm~\cite{Robins11} to identify all local maxima and compute an initial segmentation.
More specifically, by interpreting the histogram densities as function values on a quadrilateral grid, each vertex computes its steepest ascending neighbor and breaks ties through standard simulation of simplicity.
Vertices with no ascending neighbors are labeled as maxima, and the corresponding segment is defined by its {\it stable manifold} - the collection of vertices whose steepest gradient path will end at the given maximum.
Finally, saddles are identified as the highest vertices whose immediate neighbors belong to more than one segment. 
As apparent in many of the images, this approach suffers from some mesh imprinting, especially in areas of low overall density, which could likely be improved with more advanced techniques, such as the accurate Morse-Smale complexes of Gyulassy et al.~\cite{gyulassy12}.
However, as discussed in more detail below, the areas of low density in the joint histogram by definition have minimal effect on the final image as very few voxels are involved.
\rvssec{In regions of high density, we found the steepest gradient algorithm quite reliable even when dealing with various sources of noise.}
Note also that the algorithm described above does not identify the full Morse complex as defined in traditional discrete Morse theory~\cite{forman98, forman02}.
For example, we avoid the complexity of identifying saddles as edges between vertices, do not explicity represent minima or the potentially degenerate one manifolds from saddle to minima, and will not detect {\it strangulations} (saddles connected to the same maximum twice).
However, for a simple peak-finding approach, the above algorithm is equivalent to more sophisticated solutions and trivial to implement. 

\subsubsection{\rvs{Problem With Persistence Simplification }}
The next challenge is that discrete approaches like the one described above are well known to identify a large number of spurious maxima that must be discarded for the segmentation to become meaningful.
The most common approach to simplify Morse complexes is driven by {\it persistence} as proposed in Edelsbrunner et al.~\cite{edelsbrunner00} for histograms.
Other approaches include weighted sums of persistence and spatial distances~\cite{Reininghaus11tvcgb} or volumes of segments~\cite{Carr04vis}. 
The persistence in this context can be thought of as the height of a peak, i.e., the difference in density between the maximum and the highest neighboring saddle.
To simplify the segmentation, one iteratively identifies the remaining maximum with lowest persistence and merges the corresponding segment with the segment on the other side of the corresponding saddle.
However, although intuitive for terrains, persistence is not an ideal metric for bivariate histograms.
By definition, the height in a histogram indicates the number of voxels in the original data that share the corresponding value pair.
Consequently, how large a particular peak is does not necessarily correlate with its importance, but only with the quantity of the material present.
In fact, by far the largest peak is typically associated with the background material, air in our case.
Therefore, using the absolute height and even the absolute difference in height between a peak and its saddle as a measure of \rvssec{importance} will significantly overvalue common materials and ignore rare combinations, i.e., small inclusions.
Such a result is the opposite of the desired effect since the most common materials in an object are often the least interesting.

\subsubsection{\rvs{Modification for Our Cases: The Relevance Metric }}
We propose a different metric based on relative persistence, also called {\it relevance}~\cite{Mascarenhas09,Bremer16camcs} Figure\autoref{fig:persist_relevance:relevance}.
Consider two neighboring segments \rvs{ capped by maxima $m_1$ and $m_2$ separated by the saddle $s$ with $f(m_1) < f(m_2)$.
The persistence of the $m_1-s$ pair is defined as $f(m_1) - f(s)$.
Its relevance is defined as  $1 - \left(f(m_1) - f(s)\right) / f(m_2)$ as the global minimum of our histograms is assumed to be $0$.}
\rvs{Conceptually, relevance acts more like a local signal to noise \rvssec{ratio}. 
In a high-density region of the histogram, even a comparatively small amount of noise may cause large density variations. 
On the contrary, in a low-density region the same amount of noise will cause much smaller variations.
Persistence measures absolute variation whereas relevance measures local variation.
As a result, simplifying according to relevance better preserves low peaks corresponding \rvssec{to} the rare materials.
}
\autoref{fig:persist_relevance_overview} illustrates the differences between a persistence- and a relevance-based segmentation for a 1D example.
We also impose a small absolute lower bound on the density and will not consider maxima with less than a certain number of samples.
This action is necessary to prevent numerical noise from being amplified in the relevance computation in regions of minimal density. 
The final result is a simple-to-implement and fully automatic hierarchical segmentation, which identifies materials according to how (relatively) distinct they appear in the joint histogram.
\rvssec{
The result corresponds well to how our experts would manually perform the segmentation and avoids common problems when colormapping histograms of large ranges of densities.}

We choose relevance because favoring smaller features instead of further subdividing higher peak gives a better segmentation for our datasets. 

\section{Histogram-Based \rvssec{Bimodal} Renderer}\label{section_renderer}

Based on the segments from the histogram segmentation, we can now render the two volumes. We developed a \rvssec{bimodal} renderer that loads the segmentation as 2D texture. 
%Our visualization system consists of two parts: first part computes the hierarchical topological segmentation, a distance field, and a OSPRay-powered multi-modal renderer, that loads the segmentation and distance fields as 2D textures. 
%a light-weighted Python program to generate topological segmentation and a distance field as png images, and a renderer based on the ray tracing engine OSPRay. 
\rvssec{
Together with the segmentation images, the \rvssec{bimodal} CT datasets are passed to the renderer as raw format. 
%As mentioned in the previous section, all the histogram segmentation computation was done in full resolution, and a multiresolution volume loading mechanism from OpenVisus is applied to larger volumes to obtain a downsampled one that then fits into the memory and guarantees an interactive rendering refreshrates. 
}To reach the goal of a joint view of both X-ray and neutrons \rvs{CT images} for domain analysis, the two channels are composited in the renderer in a user-specified configuration. The options cover a wide range of operations, from choosing the number of segments to setting the color or visibility of individual segments. In this section, we describe the details of our implementation. \rvssec{The entire project is open-source and can be found at: https://github.com/xuanhuang1/multiChannelOSPModule}.

\subsection{Co-Registered X-Ray and Neutron Volume Renderer}
We used the open-source ray tracing engine OSPRay\cite{Wald16}, and modified the built-in Scivis renderer for a true \rvssec{bimodal} ray-casting rendering pipeline. \rvssec{Combined with the histogram segmentation, our visualization encodes original information from both channels.}
%Only one volume is traced and is sampled for both channels. We are able to To achieve better rendering results on common desktops.
%This single-pass tracing enables us not only to show the correct depth of the multimodal dataset, but also to keep a higher frame rate compared to traditional methods that combine two rendering passes in a postprocess manner.

The transfer function is defined as a color lookup from the segmentation images. For each voxel sampled, the density value pair from the X-ray and neutron \rvs{data} determines its 2D coordinate in the color segmentation image, which is passed to the renderer as a 2D texture. Then the corresponding pixel on the texture is sampled, and its RGB color is assigned to the voxel. Opacity is set as uniform value for all segments. %\ptb{Is this supposed to say the entire *segment*?} \rvs{It's a constant opacity for all segments, so yeah the entire dataset.}
%Opacity is determined in a similar way with the same 2D coordinate. Instead of looking up the color segmentation image, the opacity lookup is done with the gray-scale distance field image. A distance field to opacity function is applied on top of the scalar value obtained to produce the final voxel opacity.
The resulting renderer is a modified ray-caster with a specialized transfer function mapping. We chose our current pipeline for interactivity and a straightforward rendering result to users. %\rvssec{With us building on an open-source rendering engine, if further improvements on image quality is desired, advanced rendering techniques such as denoising or path tracing can also be integrated conveniently based on our application as well as the OSPRay library itself}. 

\subsection{Interactions}\label{section_interaction}
%\rvs{

\begin{figure}[htb]
\centering
\subfloat[][Main panel]{
  \includegraphics[width=0.48\linewidth]{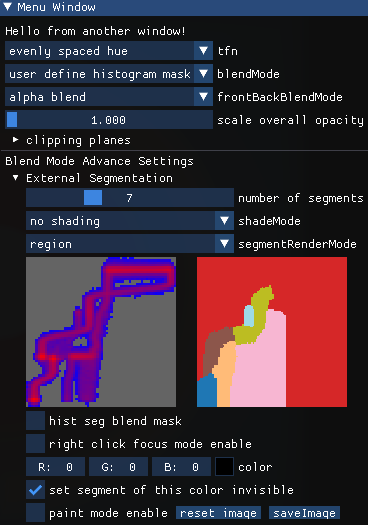}}
  \subfloat[][Histogram mask option and paint mode enabled]{
  \includegraphics[width=0.48\linewidth]{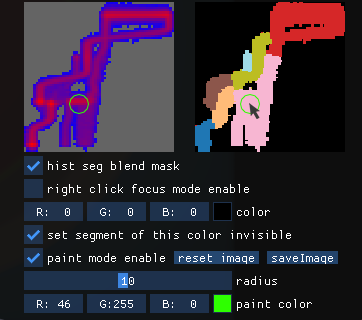}}
  \caption{\label{fig:gui}\rvs{Graphics user interface. The main menu in (a) lists user operations to set shading options and the number of segments. By clicking on the segmentation image, the user is able to set color and visibility for each segment. As shown in (b), a histogram over segmentation mask option is available for visual correlation. A paint mode widget (enabled here on the right) is \rvssec{located} at the bottom for further manual segmentation editing.}
  Extra utilities include global opacity scalar and clipping planes. The entire interface consists of only commonly seen elements and simple operations. 
  }
\end{figure}
%}
%\rvs{
In order to reduce the amount of interactions, the segmentation process only requires the user to select the number of segments. When the volumes do not contain any noise, the number of segments correspond exactly to the number of materials plus the background. However, as X-ray and neutron \rvs{scans} will not necessarily capture all materials as the same density value pairs, as described in \autoref{sec:morsecomplex}, the number of segments will also depend on the quality of the scans, the material, and the location in the volume. By providing the flexibility to adjust the number of segments, we allow the user to explore different segmentations without the need to  manually draw polygons in the histogram, or segment \rvssec{the} volume. The \rvssec{bimodal} renderer provides a quick overview of the segmentation, as shown in \autoref{fig:overview}. \rvs{The main features of interaction are listed below:}

\rvs{
\begin{itemize}
\item
To resolve \rvssec{the} inherent visual occlusion problem with volume visualization, we provide the user with two ways to toggle the visibility for all available segments: all visible except for the selected segment or none visible except for the selected segment.
\item 
Further flexibility and support for exploration is provided through the color assignment for each segment. 
\item An overall opacity scalar value slider for quick transparent overview or solid material observation.
\item A segmentation editing panel that allows free color painting directly on the histogram segmentation result, acting as a manual refinement tool on top of an automatic segmentation. This widget interactively updates the volume rendering mapping, which enables user interaction with small 2D areas in bivariate space plus the the precision and convenience of an automated process. \autoref{fig:gui} shows individual components in detail.
\item
Since the 2D space pixel color is simply the segmentation ID for each material, the resulting image from editing encodes the desired segmentation for this visualization session and can then be saved as a simple png file for further analysis tasks. 
\end{itemize}
}
%Further flexibility and support for exploration is provided through the color assignment for each segment. Although the initial color is preassigned via a qualitative color map, the user can reassign colors. After clicking on the segmentation image, the user can change the color with a color picker and toggle the visibility.
%For example, the user may assign the same color to different segments in case they correspond to the same material, but due to scanning artifacts, these two segments are represented by different density pairs. The overall opacity can be adjusted with a scalar value slider. If a less occluded view is desired, the user can simply set the volume to be more transparent, whereas a solid rendering is more helpful for individual material observation. 
%Finally, there is a segment editing functionality the allow the user to paint directly on the segmentation color image, which is interactively updated to the volume rendering mapping. This functionality offers another aspect of exploratory experience with the histogram, and the modified segmentation is primarily used as a manual refinement tool on top of an automatic segmentation. The user can then take advantage of the precision and convenience of automated process, as well as the ability to interact with small pieces in bivariate space. The resulting image encodes the desired segmentation for this visualization session and can then be saved as a simple png file for potentially further analysis tasks.

\section{Results and Expert Feedback}\label{section_result}

We demonstrate our approach on four datasets that our domain scientists use. From the weekly meetings with them, we derived the main case studies that we describe in the context of each individual dataset.
We collected feedback from the four domain experts by demonstrating the visualization at regular meetings and, in addition, two of the experts used our system directly to study the datasets described in \autoref{tab:datasets}\rvssec{.}
\rvs{All domain experts are senior scientists that are focused on nondestructive evaluation with more than 10 years of experience. They lead groups or projects for the characterization of materials in advanced manufacturing.}
From these informal sessions, we collected the experts' informal feedback and formed the following four case studies. We also describe our observations on their usage, what they found easy to understand and what was difficult. We summarize their responses at the end of all user case analyses in \autoref{section_feedback}.
\rvs{The users tested the approach on a laptop with an Intel i9-1088H @ 2.4 GHz processor, 64 GB RAM, and a Nvidia Quadro RTX 3000 (6 GB RAM). The application maintained at least around 10 fps on all case studies.} 

\subsection{Case Study 1 JH2B and XR05A: \rvssec{Semiautomated Fast and High-Precision Material Segmentation}}

\rvssec{In this section, we perform a qualitative, material-to-material analysis to demonstrate the effectiveness of the segmentation system. We present our result with two datasets, JH2B and XR05A. The segmentation method is also compared to the cluster-based k-means segmentation as well as Wang \etal's gradient magnitude based approach.}

 \begin{figure}[htb]
\centering
    \subfloat[][Ground truth]{\includegraphics[width=0.24\columnwidth]{img/uc1_jh2own_groundtruth.png}}\hfill
    \subfloat[][Neutron]{\includegraphics[width=0.24\columnwidth]{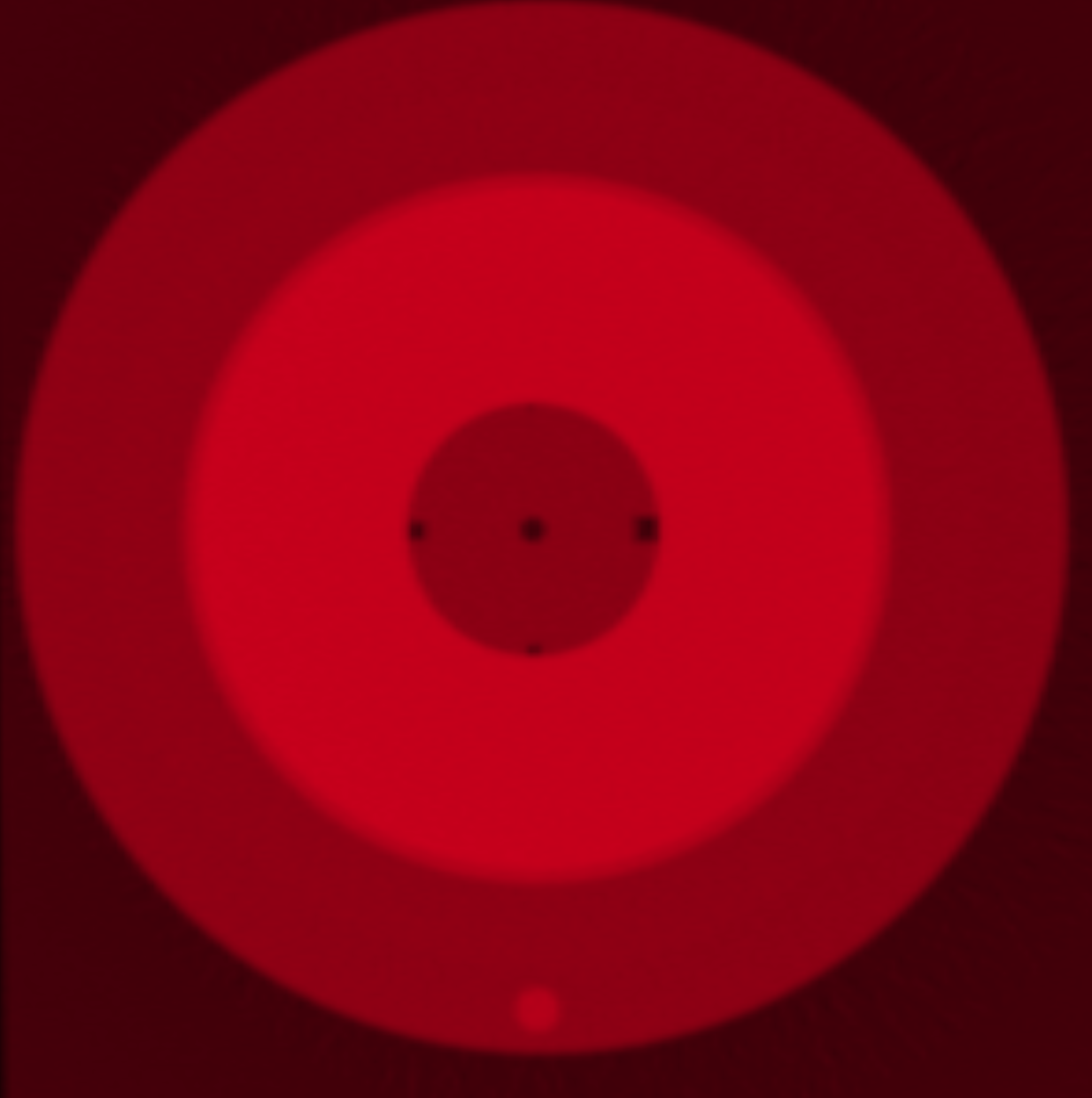}}\hfill
    \subfloat[][X-ray]{\includegraphics[width=0.24\columnwidth]{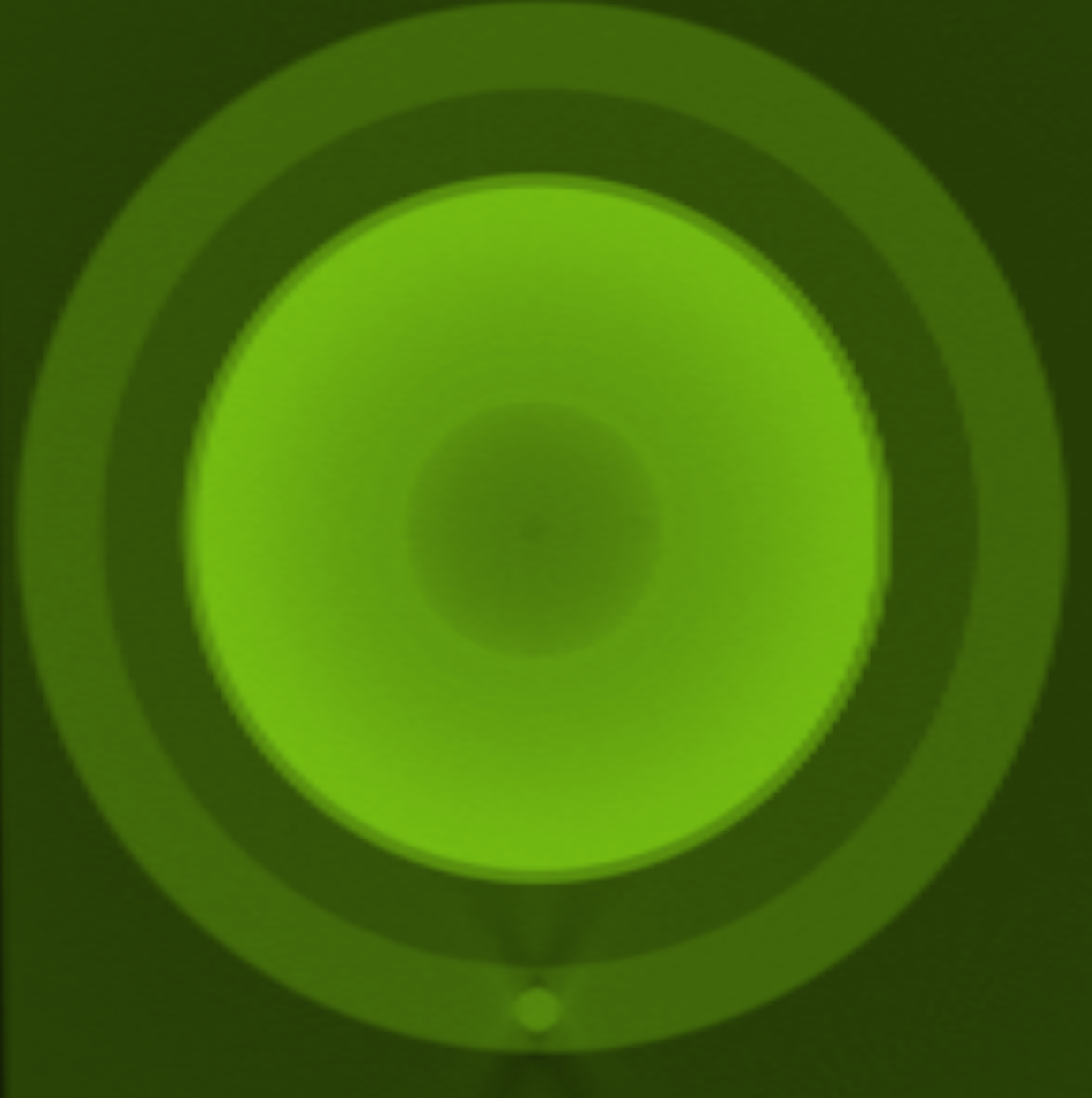}}\hfill
    \subfloat[][Our view]{\includegraphics[width=0.24\columnwidth]{img/uc1_jh2own_mmvis_modified.png}}
  \caption{JH2B ground truth, neutron (red), X-ray (green), and the \rvssec{bimodal} rendered result. The same slice of JH2B is shown. The structure consists of aluminum (orange), HDPE outside (pink), stainless steel (purple), tungsten (green), and HDPE inside (brown). Note that tungsten not only blocks X-ray from depicting the inside of HDPE, but also results in a wide range of X-ray values due to absorption. Neutron preserves the interior rods but does not distinguish aluminum and HDPE. Our final view successfully combines the desired features from two modalities. }
  \label{fig:jh2own}
 \end{figure}

\begin{figure}[htb]
\centering
    \includegraphics[width=0.24\columnwidth]{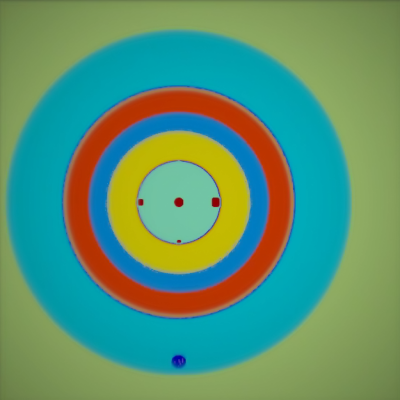}\hfill
    \includegraphics[width=0.24\columnwidth]{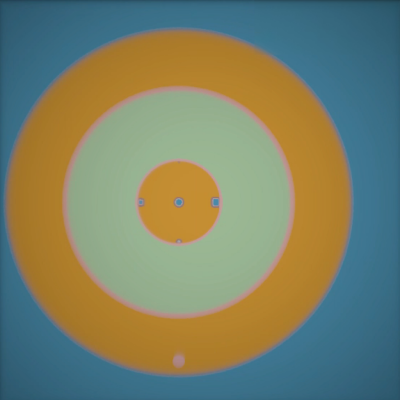}\hspace{-0.5mm}
    \includegraphics[width=0.24\columnwidth]{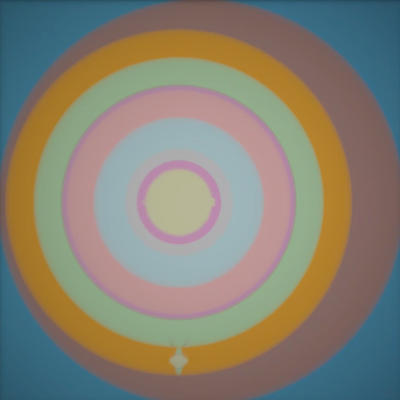}\hfill
    \includegraphics[width=0.24\columnwidth]{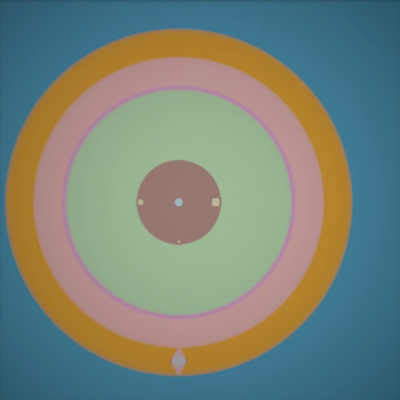}
    
    \subfloat[][K-means]{\includegraphics[width=0.24\columnwidth]{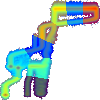}}\hfill
    \subfloat[][Wang's Neutron]{\includegraphics[width=0.24\columnwidth]{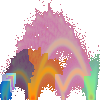}}
    \hspace{-0.5mm}
    \subfloat[][Wang's X-ray]{\includegraphics[width=0.24\columnwidth]{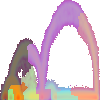}}\hfill
    \subfloat[][Ours]{\includegraphics[width=0.24\columnwidth]{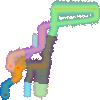}}
  \caption{ \rvssec{JH2B 2D histogram segmentation methods comparison. Column (a) shows the k-means clustering with the histogram as scatter plot, which doesn't conform to material shapes at boundaries. (b) and (c) Shows Wang \etal's method that uses gradient magnitude as the second axis on histogram for each modality and apply the Morse-complex segmentation individually. Note how the Morse-complex segmentation does not pick up individual arcs easily, and reflects the drawback of each modality as seen in Figure \ref{fig:jh2own} (b) and (c). Column (d) shows a more desired result from our method.}}
  \label{fig:jh2own_comp}
 \end{figure} 
 
In the JH2B dataset, the number of materials in the structure is known, but the number of X-ray and neutron value pairs for a material may be different depending on the other materials in the object due to the scanning artifacts described earlier. In particular, this dataset has long, thin cylindrical rods in the center of the object that often require closer attention. Here, we first demonstrate the histogram-based approach to identify material peaks. As shown in~\autoref{fig:manual }, the bivariate histogram already indicates several high peaks that our experts would associate with the corresponding materials. They also confirmed that they would draw a similar segmentation manually, but Morse-complex segmentation is able to identify a detailed boundary that would require the placement of polygons with many sides, which would be tedious to do by hand.   

The expert is interested in a closer look at the dataset to see material boundaries and any small features \rvs{in the volume}, and therefore the volume is clamped onto the slice-like view that experts are familiar with. We will also compare our result with the phantom images produced by the existing traditional slice-view method.

As we can see in \autoref{fig:jh2own}, the X-ray (green) fades undesirably with a lower precision at the center but is of higher quality overall, whereas the neutron (red) remains consistent within the materials but has lower \rvs{accuracy} in terms of reconstruction. Our result includes both the holes on the inner ring from the neutron and high-contrast boundaries from the X-ray, generating a result very similar to the ground truth. 

\rvssec{To demonstrate the effect of different segmentation methods we applied three 2D histogram segmentation on the JH2B dataset expecting nine resulting segments, as shown in \autoref{fig:jh2own_comp}. Comparing to the ground truth illustration Figure \ref{fig:jh2own} (a), the clustering-based method groups the 2D points by distance and does not catch precise shape boundaries as in the topological-based method, where the segment boundary are intentionally defined as along the local minima. 
}
% Besides the key differences in our approach that utilizes relevance-based metric and semi-automated segmentation editing, their method differ by taking only one modality as input instead of two. \ref{fig:jh2own_comp} (b) and (c) demonstrates the application of this method to our data and the resulting arcs in the bivariate histograms are not segmented as joint segments, which would not work for our data.  

\rvssec{We also compare our segmentation to the method of Wang et al. \cite{wang12}, which performs the Morse-complex segmentation on the bivariate histogram of the value and gradient magnitude \ref{fig:jh2own_comp} column (b) and (c).
whereas their Morse-complex based method segments the histogram more toward distinct peaks, the material identification of these composed histograms mainly relies on recognizing arc motifs that correspond to material boundaries. This misalignment between the nature of the Morse-complex based method and the segmentation goal results in inefficient material identification. For example, in column (b) on the bottom we can clearly see four distinct curves, but the orange and pink colors actually split and merge part of these structure randomly.} \rvssec{Besides, our key goal was to provide a joint view of the two modalities, and using the their method would still require us to mentally assembling two complementary visualizations that each lacks different pieces of information, as seen in Figure \ref{fig:jh2own} (b) and (c)). 
%Besides, for bivariate data Wang's method still requires mentally assembling two complementary visualizations that each lacks different pieces of information, as seen in Figure \ref{fig:jh2own} (b) and (c)). 
Our method in Figure \ref{fig:jh2own_comp} column (d) demonstrates a more efficient 2D histogram segmentation that prioritize key materials and is straightforward to interpret. 
}

\begin{figure}[!htb]
  \centering
  \subfloat[][Histogram]{
  \includegraphics[width=0.3\linewidth]{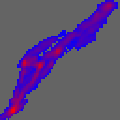}}
  \subfloat[][Segmentation]{
  \includegraphics[width=0.3\linewidth]{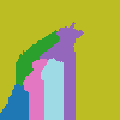}}
  \subfloat[][Segmentation with non empty points]{
  \includegraphics[width=0.3\linewidth]{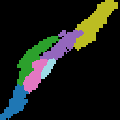}}
  \\
  \subfloat[][Modify automatic segmentation by segment merging ]{
  \includegraphics[width=0.96\linewidth]{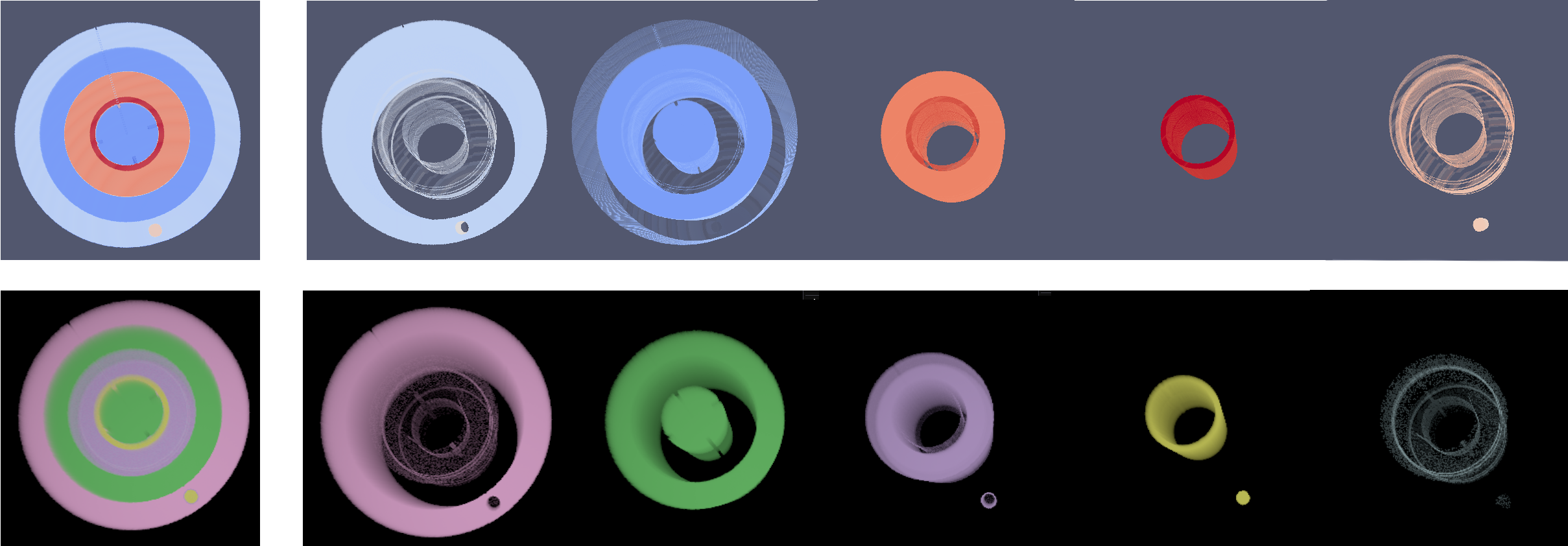}}
  \caption{\label{fig:xr05_components} \rvs{A five-component segmentation with XR05 (background set to transparent). This is a slightly different variation of the JH2B dataset. (a) Shows the histogram segmentation. The top row in \rvssec{(d)} is the manually labeled volume, which can be viewed as the ground truth of the segmentation. The second row in (d) is our result. We have reached a close approximation with only five segments. The small outer rod piece gets blended into other segments due to the value inaccuracy in the volume data itself and thus shows no distinct peak in the histogram, but can be easily identified spatially. }
  }
  \vspace{-2mm}
\end{figure}

The XR05A is another simulated dataset with a higher quality X-ray. This dataset is expected to produce high-quality segmentation, and we validate our results with a labeled volume that describes the desired segmentation.( \autoref{fig:xr05_components}). 

We are aware that the segmentation does not group components precisely, in that we can see the tube on the side appears in two segments. \rvs{Not all material is segmented into its own segment because of the inconsistency of the X-ray, resulting from the data \rvssec{acquisition} step itself and thus introducing noise into the bivariate histogram compared to the ground truth.} However, the final visual result could be improved by oversegmenting and merging, as shown in the next case study.

\subsection{Case Study 2 JH2B: \rvs{Exploratory} Material Identification via Segmentation Adjustment}
 
\begin{figure}[!htb]
\centering
  \subfloat[][Modify automatic segmentation by segment merging ]{\includegraphics[width=0.9\columnwidth]{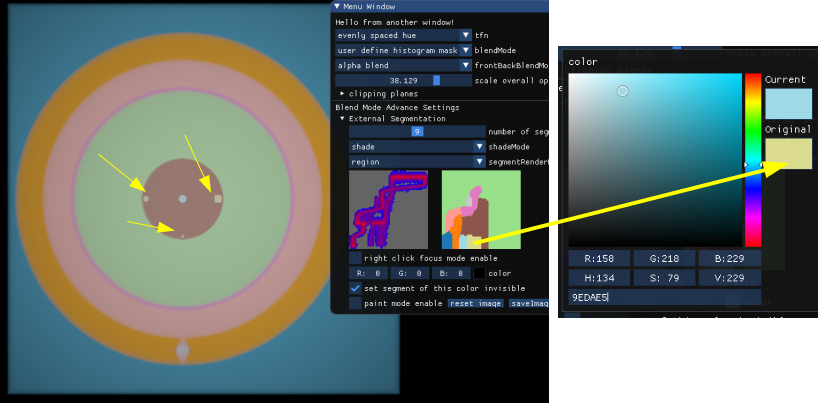}\label{fig:jh2own_interface:0}}\\
  \subfloat[][Ground truth
  ]{\includegraphics[width=0.25\columnwidth]{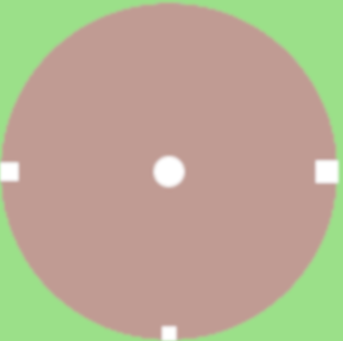}\label{fig:jh2own_interface:a}} \hspace{3mm}
  \subfloat[][Automated segmentation
  ]{\includegraphics[width=0.25\columnwidth]{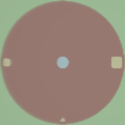}\label{fig:jh2own_interface:b}} \hspace{3mm}
  \subfloat[][User adjusted]{\includegraphics[width=0.25\columnwidth]{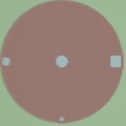}\label{fig:jh2own_interface:c}}
  \caption{JH2B segment merging: (a) shows the process of changing color with the color picker from yellow to light blue, so that the two segments get merged, and (b)-(d) are comparisons to the \rvssec{ground truth}.}
  \label{fig:jh2own_interface}
  \vspace{-5mm}
\end{figure}

One of the main goals of the visualization task is to quickly perform exploratory analysis of the structure to see if important features can be easily spotted. These exploratory tasks are achieved by adjusting the number of segments and setting the opacity. In this case study, we first demonstrate the interactive adjustment of the segmentation results with the JH2B dataset.

%\rvs{
The users \rvs{in this experiment} first applied clipping planes to cut along the xy-plane. Once a location of interest was reached, the number of segments was set to a higher number for precise segmentation. As shown in Figure \autoref{fig:jh2own_interface:b}, the histogram is initially oversegmented until all parts receive their respective segments. Here, the user decided to merge the light blue and light green segments by assigning them the same color, since they are made of the same material. 
This case study also involves the greatest amount of interactions, but the process is still intuitive as all the widgets are widely used in common scalar field visualization interfaces.
We demonstrate here the ability of the interactions to easily correct the segmentation. The result after adjustment is shown in Figure\autoref{fig:jh2own_interface:b}, which can be compared with the ground truth in \autoref{fig:jh2own}.

\begin{figure}[htb]
\centering
    \subfloat[][Unpainted]{\includegraphics[width=0.48\columnwidth]{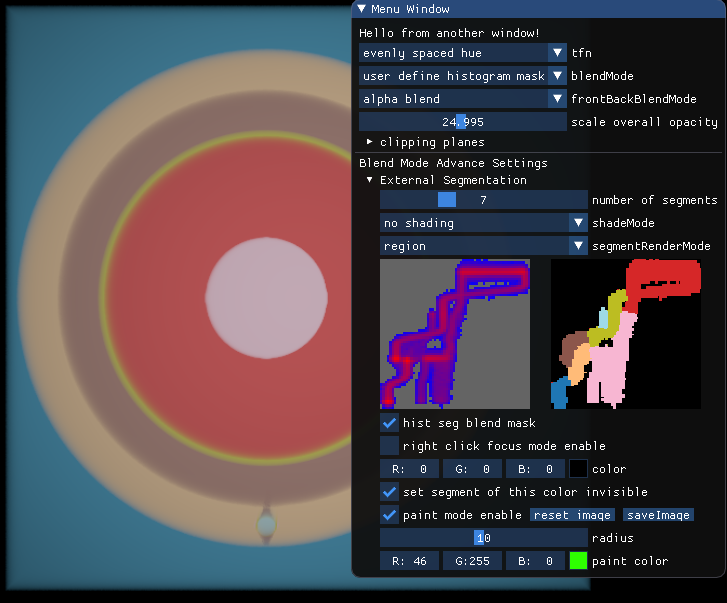}}
    \hfill
    \subfloat[][Painted]{\includegraphics[width=0.48\columnwidth]{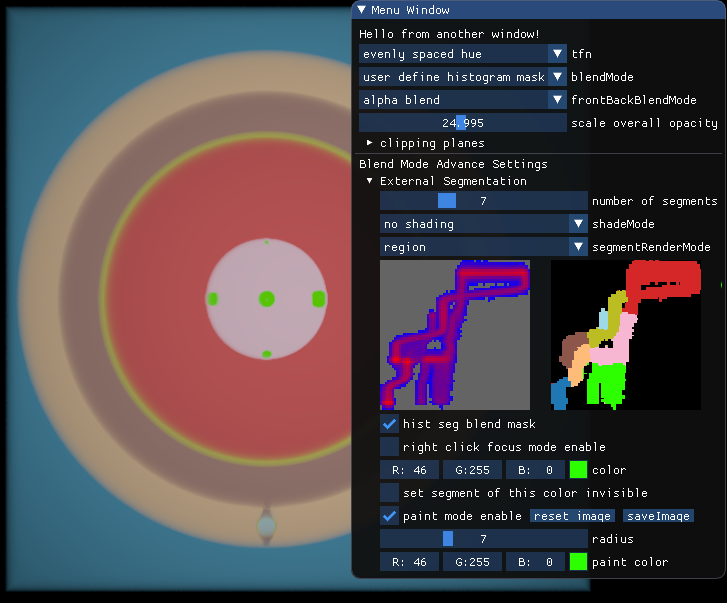}}\\
    \subfloat[][unpainted]{\includegraphics[width=0.24\columnwidth]{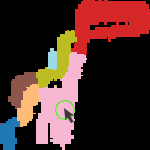}}\hfill
    \subfloat[][part 1]{\includegraphics[width=0.24\columnwidth]{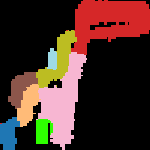}}\hfill
    \subfloat[][part 2]{\includegraphics[width=0.24\columnwidth]{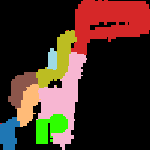}}\hfill
    \subfloat[][painted ]{\includegraphics[width=0.24\columnwidth]{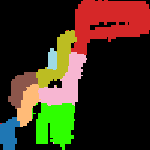}}\\
    \subfloat[][unpainted ]{\includegraphics[width=0.24\columnwidth]{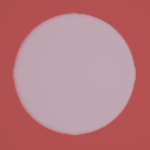}}\hfill
    \subfloat[][part  1]{\includegraphics[width=0.24\columnwidth]{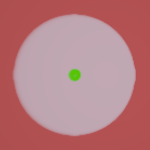}}\hfill
    \subfloat[][part 2]{\includegraphics[width=0.24\columnwidth]{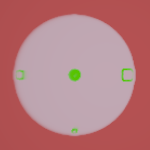}}\hfill
    \subfloat[][painted ]{\includegraphics[width=0.24\columnwidth]{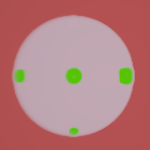}}
  \caption{\rvs{JH2B user-adjusted.} Here we show an example of an automated segmentation with manual modification. The process of painting is shown in the two bottom rows.}
  \label{fig:jh2own_paint}
 \end{figure} 

An alternative to oversegmentation with merging is to add the missing material on the histogram with a smaller number of segments. As shown in \autoref{fig:jh2own_paint}, with the case of seven segments, the structure of the outer layers is already clear, but the interior holes are still missing. \rvssec{ By looking at the bivariate histogram, the domain scientists suspect that the finger-shape areas in the bottom center likely form another distinct peak and thus represent a new material, but have not yet been segmented out from its pink surroundings. }Therefore, in this situation, the user can utilize the painting tool to test the assumption. When the paint mode is enabled, the user selected a color to paint with mouse at the desired region, and the volume was updated interactively as seen with column pairs (c)-(j). \autoref{fig:jh2own_paint}.

%\subsection{Case Study 3 XR05A: Material Segmentation Validation}

\subsection{Case Study 3 XR05B and Battery. Bivariate Data Validation with Manufactured Scanning Objects}
In this section, we present the ability to perform fast segment-based visualization on two real-wold scanning datasets, XR05B and Battery. For these two real-world datasets, although the scientists are familiar with the structure of the original constructed object, the scanning and registration quality remains unclear, and thus our work fits well with the need for a quick check of previous steps. We compare the process with the slice-based views that are commonly  used, which is also the only easily accessible validation tool that we are aware of for the bivariate datasets. 

\begin{figure}[htb]
\centering
\subfloat[][XR05B segmentation]{\includegraphics[width=0.96\columnwidth]{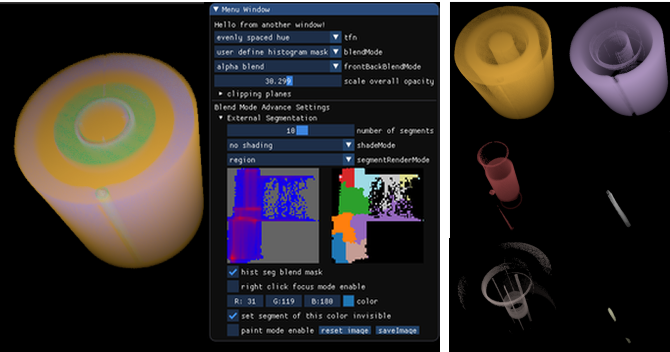}}\\
  \caption{XR05B with the collection of segments on the right. Segmentation results in partially missing pieces.}
  \label{fig:xr05B}
 \end{figure}

The \rvssec{XR05B} object has been deliberately constructed to stress the scanning process, and therefore the user of this case would like to have a fast validation of the data quality. Because the scientists rely on slicing for general initial observation, the XR05B testing object is designed to be identical along z-axis, similar to the JH2s. Based on the knowledge of such underlying structures, the scientist was able to use the automated segmentation result to spot important visual clues of potential acquisition defects. ~\autoref{fig:xr05B}.  

 \begin{figure*}[!htb]
  \centering
  \subfloat[][Battery in slice view, the top row is from X-ray and the second row is from neutron. ]{\includegraphics[width=0.92\linewidth]{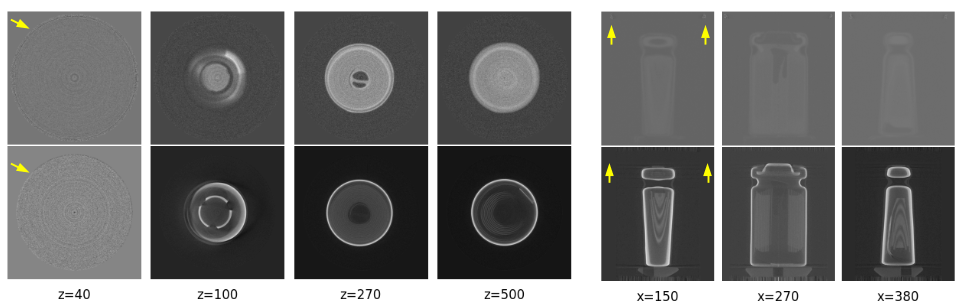}}\\

  \subfloat[][\rvs{Histogram}]{\includegraphics[height=0.12\linewidth]{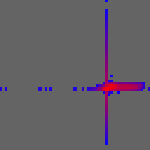}}
  \hspace{2mm}
  \subfloat[][\rvs{Segmentation}] {\includegraphics[height=0.12\linewidth]{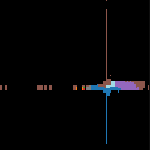}}
  \hspace{2mm}
  \subfloat[][Battery in our view. Overview and segmentation. ]{\includegraphics[height=0.12\linewidth]{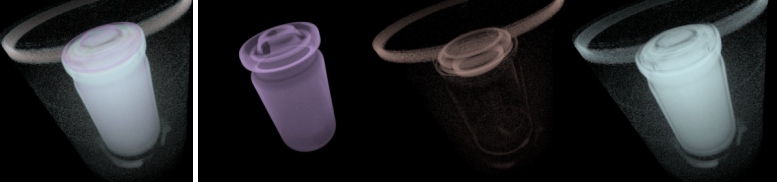}}
  \hspace{2mm}
  \subfloat[][Clamped at center ]{\includegraphics[height=0.12\linewidth]
  {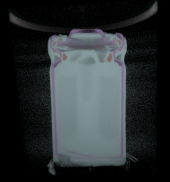}}

  \caption{Battery: (a) shows a traditional slice based view of the dataset. Our work (b) shows a more intuitive overview of the datasets, and a clamp operation as in (c) enables a close look of the segmentation similar to slicing for a familiar view. The \rvssec{bimodal} visualization system provides a good estimation of the structure through automated segmentation while avoiding dealing with a stack of 2D slices in both modalities. Note that our work also clearly captures the upper ring, which is blurry in X-ray and almost missing in neutron.}\label{fig:battery}  %\TODO{describe each segment, what's the corresponding material} 
  \vspace{-2mm}
\end{figure*}

To validate the quality of this acquired bivariate dataset, the user first noticed the bleeding effects \rvs{on the histogram (blue scattered points in ~\autoref{fig:xr05B})} spreading out straight across the axis. Those horizontal thin lines were less often seen in the actual density value pairs, but were more likely resulting from scanning or registration artifacts. Therefore, after identifying several tilted or partially missing components that corresponded to the bleeding area on the histogram ~\autoref{fig:xr05B}, the user was then able to confirm that there is a problem from a previous process, potentially a registration misalignment. This dataset would likely be unsuitable for further analysis, and the subtle misalignment would be difficult to diagnose by viewing the two channels side by side. Our system has successfully served as a fast visual system to observe the data quality resulting from the acquisition step.

For most real-world datasets, the data distribution is not as convenient as for JH2B or XR05B. Battery is another real-world scanned object that is more complicated to decipher. The traditional data interpretation process requires multiple slices at exact locations for both modalities to identify all structures, whereas with our system it is straightforward in the overview ~\autoref{fig:battery}.

\subsection{Case Study 4 Meteorite: \rvs{Investigating Distinct Material Quantities in an Unknown Object}}

%use meteorite.\TODO{description here. iron–nickel is a material of interest}

The experts \rvs{working with meteorite} had little knowledge of this dataset at the time of viewing. The object is scanned to inspect the inside materials. In this case study, the analysis is exploratory and the main goal is to obtain a comprehensive first-step view of the interior structure. The feature of interest for this dataset is the iron-nickel inclusions. The assumption is that this alloy lies in the meteoroid and can be well detected by the scanning devices, as the substance has different density values from the surrounding rocks. However, since iron–nickel is a small feature scattered across the entire object with no uniform shape, this dataset is considered an extremely labor-intensive case for traditional material identification techniques.

\begin{figure}[!htb]
  \centering
  \subfloat[][\rvs{Segmentation}]{
  \includegraphics[height=0.5\linewidth]{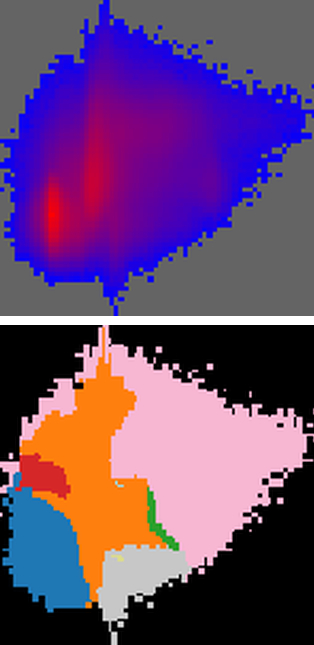}}
  \subfloat[][\rvs{Segments collection}] {
  \includegraphics[height=0.5\linewidth]{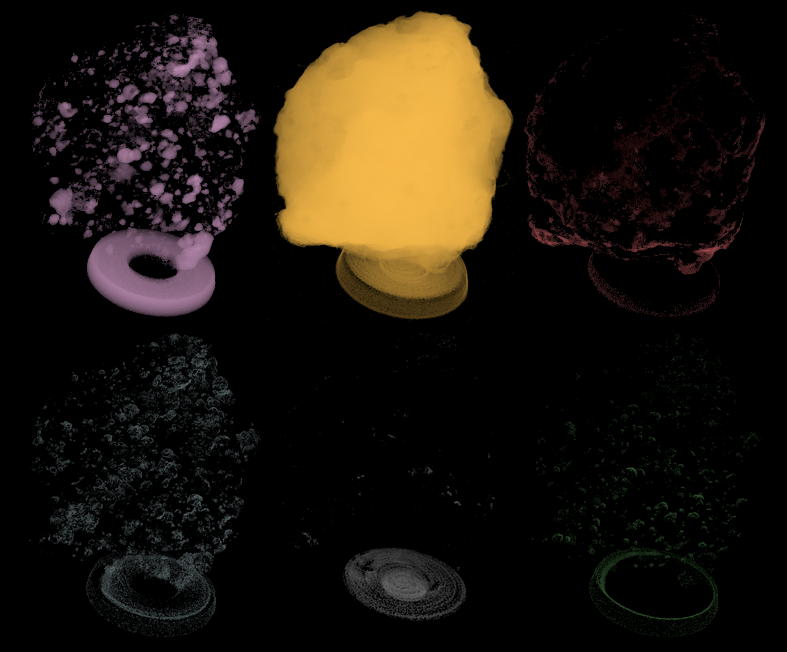}}
  \caption{\label{fig:meteorite_components} \rvs{Meteorite for case study 4}. Showing collection of individual segments. The first segment in purple corresponds to the domain experts' expectation of the iron–nickel alloy. The rest are different parts of rock. }
\end{figure}

% \begin{figure}[htb]
%   \centering
%   \includegraphics[width=0.96\linewidth]{img/meteorite_3.png}
%   \includegraphics[width=0.96\linewidth]{img/meteorite_10.png}
%   \caption{\label{fig:meteorite_segs} Different number of segments on meteorite dataset}
% \end{figure}
With our visualization system, the analysis started by removing the background segment. After changing the number of segments a few times with a slider, the user observed that the majority of this object is rock, and thus further segmentation is not necessary. When satisfied with the segmentation, the user was then able to loop through each individual segment easily and got a good initial understanding of the dataset itself. 
\rvs{The users stated that this is a highly relevant use case for our method, as they can use our tool to clearly view the various inclusions in the object with just a few clicks. This tool helps to make a more informed decision about further nondestructive evaluation techniques.}
%Within just a few clicks by the user, our work clearly segments out a purple region over the histogram, which is likely the iron–nickel as indicated by the user. The user is then confident to move forward with this dataset to a more detailed analysis. A further investigation of the material can also be performed with a slicing operation similar to that in Case Study 2. 

% \rvs{
% \subsection{Interactive Material Identification}
% \begin{itemize}
%     \item battery, XR05, meteorite
%     \item info about the dataset
%     \item how long it took 
%     \item talk about instant feedback from renderering
% \end{itemize}
% }

%\begin{figure}[htb]
%  \centering
%  \includegraphics[width=.5\linewidth]{img/jh2_histogramWeighted.PNG}
%  \caption{\label{fig:JH2_histWeighted}
%           Here is a sample figure.}
%\end{figure}

\subsection{Expert Feedback}\label{section_feedback}

The four experts in nondestructive evaluation also provided qualitative feedback on the current prototype of our approach. Three experts were involved in the design of our approach and contributed valuable data, regular discussions, and testing of our system, and hence they are also included in the author list. Their expertise ranges from CT reconstruction, analysis, and evaluation, to experimental validation.

\rvssec{The fourth} expert who had previously focused on finding appropriate methods for visualization \rvssec{of} X-ray and neutron data was not involved in the design process of our approach. He provided valuable feedback, but is not included in the author list. For these four experimental sessions,  two experts were able to directly use the developed system; one tested the system remotely via screen sharing, and the other was shown the results directly and provided feedback. A third expert was able to use the manually painted segmentation image afterwards to extract surfaces for further analysis. 

We have received generally positive feedback from our target users and grouped them into four categories, as discussed below. After a short demo from us, the experts were able to understand the theoretical background of the segmentation for data interpretation. They were able to identify materials through simple view manipulation and segment color toggling.

\subsubsection{Efficient System for Material Identification}
One aspect that was praised by all experts is the ability to easily select the material regions from the bivariate histogram without the need to draw the exact material boundaries themselves. One expert had previously worked on manual segmentation of the bivariate histogram and was very impressed with the ability to automatize this process. Not only does our system automatize the drawing of polygons, but the detected boundaries can also be much more detailed than manual drawing can achieve. According to him, this manual segmentation was the most laborious part of the previous analysis procedure and our automated segmentation saves around 20 minutes of manually placing polygons for each dataset. Furthermore, the experts appreciated being able to simply change the number of segments.

The \rvssec{bimodal} visualization has also been reported to be helpful in highlighting an overview of all material boundaries, which is difficult to obtain with the traditional slicing method. However, for a detailed examination, the experts preferred to look only at smaller slabs, using clipping planes, to avoid visual occlusion issues. One expert compared our work to the fusion of X-rays and neutrons but preferred our semiautomated method to change the combined view and the number of segments. With this method, artifacts and errors in the acquisition as well as imperfect automated fusion algorithms can be adjusted for. 
\subsubsection{Self-Explanatory Concept and Familiar View}
The experts were very satisfied with the simplicity of the visualization and the incorporation of familiar views, such as the bivariate histogram and the intuitive histogram-to-volume color mapping. The visual result did not introduce any "black boxes", such as in the deep learning approach, that would decrease confidence in the interpretation of the results. Instead, we use well-established bivariate histograms and combine them with a segmentation that can be simply understood by the user (using hills and valleys). The experts immediately understood that the Morse-complex segmentation is related to \rvs{the common watershed solution, but preferred our hierarchical method due to the ability to simply adjust the number of segments. }

\subsubsection{Minimum Data Manipulation}\label{min_data_manipulation}
A senior member of the nondestructive evaluation team appreciated the fact that our method did not modify the datasets as typical data analysis techniques  do. The result from our automated segmentation is not only flexible enough to play around in real-time, but it is also not "corrected" to answer any specific question. Rather, the final combined view remains as a direct representation of the raw input. For example, the material closer to the center easily gets oversegmented due to X-ray decay. The experts commented that although it may take some thinking to see the reason behind this tendency, the result revealed the underlying physics of how the beams behave for these cylindrical objects. Thus, the visualization system itself was considered a more faithful representation of the dataset, and thus is well suited for general purpose analysis.

\subsubsection{Further Expectations}
The scientists have expressed an overall interest in using the system further with more datasets because they believe this tool can be very helpful in general. One of them stated he would like to see further integration of other data analysis toolkits to make full use of the visualization, for example, validating the boundaries that are shown in the view.

The main negative comments about the visualization come from two aspects, the quality of automated segmentation and the lack of shading. The final segmentation is not as accurate as a data-driven approach. However, experts did acknowledge that our histogram-based segmentation approach is not a 'black-box', and that the purpose of our system is not to replace specialized analysis \rvssec{tools} but to give an overview. The ability to manually adjust the segmentation results enables them to account for inaccuracies caused by artefacts. The fact that we encourage the user to oversegment first is also unintuitive at first. In the end, the scientists commented that this is something they need to get used to, as the number of material peaks in the histogram does not correspond to the number of actual materials, but  \rvssec{different density} clusters in the data.

%During the automated segmentation, certain materials split into too many segments because of the value pair difference within each material. This behavior results from the artifacts of X-ray at data acquisition where the X-ray scan decays when penetrating those materials. Although this redundant segmentation is a faithful representation of the input data, it often contributes little to scientific analysis and thus could be visually distracting. However, as we have demonstrated in Use Case 3, the user will be able to correct the oversegmentation by \rvs{merging the segments with a few clicks in the proposed interface.}
%assigning segments the same color. 
The complaints of no shading come from an expectation of creating underlying geometries. \rvssec{For instance, the scientists often have some knowledge of the object and are aware that there may be surface-like materials to be extracted for further analysis, which is beyond the topic of visual representation here.} The segmented volumes are also not always clean enough to form crisp geometry structures like surfaces, and it would be time consuming to compute all the potential geometry shapes in 3D space with different number of segments. The process could be a one-time computation but will defeat the purpose of our method as being fast to use. The experts wanted surface-like shading with thinner materials of high opacity to \rvssec{enhance their} spatial perception \rvssec{of such structures}. It is possible to create a customized surface segment and extract the corresponding data points later on, but the quality remains unclear since the segment is manually defined in histogram space.

%One unintuitive aspect in user interaction here is that we encourage users to oversegment the histogram first and then merge the segments afterwards. Such an expectation was, at first, slightly confusing, as the user wanted to adjust the number of segments to the exact number of materials. This limitation comes from the quality of the segmentation itself \rvs{that is typically sensitive to the noise in the data}. Since there is no universal solution to always reach the desired segmentation with an exact number of materials, as expected from a manual procedure, oversegmentation serves as an approximation to the optimal solution by subdividing some of the components more than necessary. We consider this process as only a small inconvenience in the user experience because it is also easy for the user to identify and recover the subdivided materials into a complete segment. In the end, the scientists commented that this oversegmentation is something they need to get used to, as the number of material peaks in the histogram does not correspond to the number of \rvs{actual materials, but the densities clusters in the data}. The assignment of color and opacity was regarded as straightforward and simple in comparison to the existing approach to edit transfer functions in detail. 

%\subsection{Image Quality}
%Advantage in visualization result, comparing to existing slices. 
%\subsection{Automated Segmentation}
%Show results of different number of segmentation comparing to ground %truth.

\rvs{\section{Discussion and Limitations}}
After the system testing sessions and discussions, the experts confirmed that this research is already showing great potential to characterize multimaterial industrial objects. Although our approach is also ready to be applied to X-ray and ultrasound data, one key challenge is that these two modalities must be registered first, and we seek to extend this work to a wider range of multivariate volume datasets. In the future, we also aim to explore the possibility to edit and correct the segmentation, allowing users full flexibility to create \rvssec{a} segmentation that are ground truth quality. 

Our visualization currently does not support \rvs{integrated} multiresolution data loading. \rvs{The automated multiresolution loading process will open up the possibility of straightforward large-scale data analysis}. The segmentation quality can also be improved by adopting ideas from higher dimensions and calculating high-dimension distances \rvs{for more complicated underlying topological structures}. Other distance measures (geodesic distance for example) and more advanced topological algorithms should also be taken into consideration.

Besides direct volume rendering, it will also be preferable to enable mesh geometry extraction and surface rendering. We would also like to dig further to adapt the properties of X-ray and neutron for precise material identification, \rvssec{ for instance to favor the high-resolution X-ray before it hits any blocking material, and bias toward the neutron afterwards}. \rvs{Such physics-based} adjustments can be applied to aid further data analysis, and with further modification, the system can be easily adapted to other types of \rvssec{bimodal} medical data.

%\subsection{User feedback}
%user feedback
\rvs{\section{Conclusion}}
With the ability to rapidly inspect a joint view of X-ray and neutron data, our collaborating expert scientists can now quickly investigate the material composition of industrial objects. This additional visualization capability is essential for the development of advanced material characterization in nondestructive evaluation. We have successfully developed a \rvssec{bimodal} visualization system that provides an efficient overview of both modalities and is easy to navigate. By incorporating the well-known bivariate histogram, we ensure that the proposed approach can be easily understood by expert scientists, which is confirmed by expert feedback.

%The provided interactions for the segmented histogram allow for easy color and opacity mapping without the need to set complex transfer functions for multiple volumes. 
%Although traditional volume rendering is considered a solved challenge by the scientific visualization community, the advent of multiple-field data creates a new family of problems that cannot be solved by traditional volume rendering and transfer function design techniques. 
%In nondestructive evaluation, technical advances in scanning techniques are resulting in more and more data being collected for the same industrial object. 
%According to our collaborating experts, when the amount of acquired data is increased without the proper tools to inspect them, the evaluation of these objects will ultimately be limited. Our experts noted that the analysis still has to catch up with the development of acquisition techniques for nondestructive evaluation and our method constitutes crucial step forward. Therefore, there is an urgent need to visualize data from different modalities with simple yet effective interactions, which we address in this work. 

%\TODO{boundaries could be better (see Gyulassi's work)}. 
%\TODO{we rely on co-registered data}
%\TODO{apply to other types of \rvssec{bimodal} data (eg. ultrasound)}
%\TODO{potential application: medical domain xray vs MRI, or dual energy CT}

%\TODO{fture work: how can histogram segmentation be applied to other domains}
%\TODO{apply idea of topological segmentation of higher dimensional data wher eyou have colocated data points}
%\TODO{what other future work?}

\vspace{6mm}

\ifCLASSOPTIONcompsoc
  \section*{Acknowledgments}
\else
  \section*{Acknowledgment}
\fi

The authors wish to thank the reviewers as well as Jacob LaManna at NIST, and Allan Treiman at the Lunar and Planetary Institute for providing the sample datasets and the helpful feedback. 
%We thank Hyojin Kim from LLNL for letting us use his x-ray and neutron simulation framework. 
This work was supported by the US DOE LLNL-LDRD 20-SI-001 and was performed under the auspices of the U.S. Department of Energy by Lawrence Livermore National Laboratory under Contract DE-AC52-07NA27344 (LLNL-JRNL-830600). This work was also funded in part by
NSF OAC awards 2127548, 1941085, 2138811 NSF CMMI awards 1629660, DoE
award DE-FE0031880, and the Intel Graphics and Visualization Institute
of XeLLENCE, and oneAPI Center of Excellence.

\vspace{-1mm}
\begin{IEEEbiographynophoto}{Xuan Huang} is a PhD student at the Scientific Computing and Imaging Institute at the University of Utah, under the supervision of Valerio Pascucci. Her research interest involves scalable scientific data visualization systems and distributed computing of large-scale data.
\end{IEEEbiographynophoto}
\vspace{-10mm}
\begin{IEEEbiographynophoto}{Haichao Miao} is a computational scientist at the Center for Applied Scientific Computing at Lawrence Livermore National Laboratory, United States. His research focuses on multifield, virtual reality and large-scale data visualization. He received his PhD in computer science from TU Wien, Austria, in 2019 and worked in the Molecular Diagnostics Department at the Austrian Institute of Technology, where he focused on the development of in silico methods for the design and visualization of nanostructures.
\end{IEEEbiographynophoto}
\vspace{-10mm}
\begin{IEEEbiographynophoto}{Andrew Townsend} is an engineer in the Nondestructive Evaluation Group at the Lawrence Livermore National Laboratory (LLNL) in Livermore, CA. Current work includes X-ray radiography and computed tomography research, characterizing complex materials and structures. He was a quality manager in the aerospace industry for 20 years before transitioning to research. Dr. Townsend has authored over a dozen peer-reviewed journal papers. He holds a PhD in Mechanical Engineering from Huddersfield University, UK.
\end{IEEEbiographynophoto}
\vspace{-10mm}
\begin{IEEEbiographynophoto}{Kyle Champley}, PhD: Kyle joined LLNL in 2012 and is an applied mathematician who works in the Nondestructive Characterization Institute and is the lead for the Signal and Image Processing Research group.  He develops algorithms and writes software for data processing and reconstruction of Computed Tomography (CT) data.  Kyle is the primary developer for the Livermore Tomography Tools (LTT) software package.  He currently serves as an independent contractor for two medical imaging startup companies, Imatrex and PET/X.  Previously, he was a staff scientist at the General Electric Global Research center where he developed CT reconstruction algorithms for GE’s Revolution CT system.  Kyle received his PhD in Electrical Engineering from the University of Washington where he performed research in Positron Emission Tomography (PET).
\end{IEEEbiographynophoto}
\vspace{-10mm}
\begin{IEEEbiographynophoto}{Joseph Tringe} leads a dynamic group of more than 40 engineers and technicians in the Lawrence Livermore National Laboratory (LLNL) Materials Engineering Division, focused on characterizing a wide range of complex materials, structures, devices and systems.   Techniques of interest to the nondestructive evaluation (NDE) group include x-ray computed tomography (CT), laser and transducer-based ultrasound characterization, microwave interferometry and microwave interrogation of embedded sensors.  Dr. Tringe has led the group since 2017.   From 2003 to 2017, Dr. Tringe was a staff scientist in the Physics and Life Sciences Directorate where he synthesized and characterized devices and materials with novel electrical, optical and molecular transport properties.  He also developed advanced methods for characterizing phenomena associated with energetic materials safety.     Prior to joining LLNL in 2003, he led a group in the Air Force Research Laboratory (AFRL) Space Vehicles Directorate which created and characterized radiation- and defect-tolerant electronic materials, devices and circuits for satellites.   Dr. Tringe has authored more than thirty-five peer-reviewed publications and five U.S. patents.   He holds a PhD in Materials Engineering from Stanford University and a bachelor’s degree in Physics from Harvard College.
\end{IEEEbiographynophoto}
\vspace{-10mm}
\begin{IEEEbiographynophoto}{Valerio Pascucci} is the founding Director of the Center for Extreme Data Management Analysis and Visualization (CEDMAV) of the University of Utah. Valerio is also a Faculty of the Scientific Computing and Imaging Institute, a Professor of the School of Computing, University of Utah, and a Laboratory Fellow, of PNNL. Before joining the University of Utah, Valerio was the Data Analysis Group Leader of the Center for Applied Scientific Computing at Lawrence Livermore National Laboratory, and Adjunct Professor of Computer Science at the University of California Davis. Valerio's research interests include Big Data management and analytics, progressive multi-resolution techniques in scientific visualization, discrete topology, geometric compression, computer graphics, computational geometry, geometric programming, and solid modeling. Valerio is the coauthor of more than one hundred refereed journal and conference papers and has been an Associate Editor of the IEEE Transactions on Visualization and Computer Graphics. 
\end{IEEEbiographynophoto}
\vspace{-10mm}
\begin{IEEEbiographynophoto}{Peer-Timo Bremer} holds a shared appointment at Lawrence Livermore National Laboratory's (LLNL's) Center for Applied Scientific Computing (CASC), focusing on large-scale data analysis and visualization, and at the University of Utah, serving as Associate Director for Research of the Center for Extreme Data Management Analysis and Visualization (CEDMAV). His research interests include large-scale machine learning, data analysis, visualization, medical image analysis, topology, volume modeling, and virtual reality.
\end{IEEEbiographynophoto}

%\newpage



%\vfill


\end{document}